\documentclass[
 reprint,
 superscriptaddress,
 amsmath,amssymb,
pra,
]{revtex4-2}

\usepackage{physics}
\usepackage[utf8]{inputenc}

\usepackage{hyperref}
\usepackage{graphicx}
\usepackage{xspace}
\usepackage{xcolor}

\usepackage{amsmath}

\begin{document}

\title{Classical phase unwrapping enables large-time-step unitary quantum Krylov subspace diagonalization}
\author{Tushar Kumar Jain}
\affiliation{Department of Chemistry, Indiana University, Bloomington, IN 47405, USA}
\author{Rahul Maitra}
\affiliation{Department of Chemistry, Indian Institute of Technology Bombay, Powai, Mumbai 400076, India}
\author{Nicholas J. Mayhall}
\affiliation{Department of Chemistry, Indiana University, Bloomington, IN 47405, USA}

\begin{abstract}
For computing molecular energies on early fault-tolerant hardware, unitary quantum Krylov subspace diagonalization (UQKSD) avoids both the deep circuits of quantum phase estimation and the Pauli-term sampling of variational methods by extracting the energies from a projected real-time propagator. At a small time step $\tau$, nearly linearly dependent Krylov vectors make the overlap matrix ill-conditioned. Our tests at fixed total evolution time show that additional Krylov vectors then add no accuracy, while the thresholded energy becomes increasingly sensitive to the choice of cutoff. A larger time step improves the conditioning and reduces the Krylov dimension $M$ needed for a target accuracy, but the energies are then known only modulo $2\pi/\tau$. We present Classical Unwrapping for Resolving Eigenvalues (CURE-QKSD), a classical post-processing method that leaves the UQKSD circuits and generalized eigenvalue problem unchanged and recovers the missing multiples of $2\pi/\tau$ from phase-consistency landscapes of one or, when needed, two UQKSD runs. In noiseless tests on six molecules with 12--20 qubits, all 28 targeted energies are recovered in their aliased regimes with final errors of at most $1.1\times10^{-7}$~Ha. Under Gaussian noise with $\sigma=10^{-4}$, the median error of the ground or first excited state reaches chemical accuracy at large aliased time steps with $M=20$--$65$, whereas uncorrected principal-branch energies are off by $0.7$--$115$~Ha. CURE-QKSD therefore makes aliased time steps usable for UQKSD calculations with smaller, better-conditioned Krylov bases.
\end{abstract}

\maketitle

\section{Introduction}
\label{sec:intro}
Obtaining accurate ground- and excited-state energies of molecules and materials is a central problem in quantum chemistry, and one in which quantum computers are expected to outperform classical ones: the dimension of the electronic Hilbert space grows exponentially with the number of active spin orbitals, so exact diagonalization is classically intractable for all but the smallest systems. Realizing this advantage, however, requires algorithms whose qubit, circuit-depth, and measurement requirements match the hardware that is becoming available. Superconducting processors have recently operated a surface-code quantum memory below the error-correction threshold, with a logical error rate that decreases as the code distance increases~\cite{Acharya2025}. Such below-threshold operation is a prerequisite for scalable fault-tolerant computation. Quantum phase estimation (QPE) is the most direct route to molecular energies on fault-tolerant hardware, encoding the phase of a controlled real-time evolution in an ancilla register~\cite{Kitaev1995,Abrams1999,AspuruGuzik2005}. 
While this provides a path to exact energy evaluation,  QPE requires a trial state with sufficient overlap with the target eigenstate, which can be difficult to prepare for strongly correlated molecules~\cite{Fomichev2024,Halder2023}. Resolving an energy to an uncertainty $\delta E$ further requires coherent evolution to times of order $1/\delta E$, and for chemically relevant molecules the resulting error-corrected circuits are deeper than early fault-tolerant devices are expected to support~\cite{Cao2019,Lee2021,Katabarwa2024}.

The variational quantum eigensolver (VQE) was introduced to remove this coherence requirement~\cite{Peruzzo2014}. VQE prepares a parametrized trial state on the quantum device, measures its energy, and updates the parameters through a classical optimization loop, so its circuits can be far shallower than those needed for QPE. This shallow-circuit advantage shifts most of the computational burden to measurement and optimization. After a fermion-to-qubit mapping, a molecular Hamiltonian decomposes into a sum of many Pauli strings, whose expectation values must each be estimated by repeated measurement~\cite{Anurag2026}. For small organic molecules, a single VQE energy evaluation to chemical accuracy can require billions of measurements~\cite{Gonthier2022}, and this cost is repeated at every optimization step. Randomly initialized, sufficiently deep hardware-efficient ans\"{a}tze~\cite{Kandala2017} can also exhibit barren plateaus, in which the gradient variance vanishes exponentially with system size~\cite{McClean2018}. Adaptively constructed ans\"{a}tze such as ADAPT-VQE~\cite{Grimsley2019}, which grow the circuit one operator at a time based on energy gradients, mitigate this problem~\cite{Grimsley2023}. While selecting each operator requires additional gradient measurements, improved operator pools and subroutines have substantially reduced this overhead and the total measurement cost~\cite{Shkolnikov2023,Ramoa2025}. Nonetheless, energies and gradients must still be estimated by sampling at every step of the optimization loop. These limitations motivate methods that require shallower circuits than QPE and fewer measurements than VQE, without an iterative optimization loop.

Quantum Krylov subspace diagonalization (QKSD)~\cite{Motta2020,Motta2022,Motta2024} requires no iterative optimization. It builds a small, nonorthogonal basis from real-time snapshots of one or more reference states and obtains the energies from a single classical generalized eigenvalue problem (GEVP). QKSD has been demonstrated on superconducting hardware for strongly correlated lattice models with up to 56 sites~\cite{Yoshioka2025}. Its measurement cost, however, depends on which operator is projected onto the Krylov subspace. Two independent real-time formulations of QKSD, which appeared at roughly the same time, both project the Hamiltonian directly, an approach we refer to as Hamiltonian QKSD (HQKSD). Parrish and McMahon introduced quantum filter diagonalization (QFD), modeled on classical filter diagonalization~\cite{Wall1995,Mandelshtam1997}, in which time-evolved reference states form a variational basis for a Rayleigh--Ritz calculation~\cite{Parrish2019}. Stair, Huang, and Evangelista introduced the multireference selected quantum Krylov (MRSQK) method, which builds the Krylov space from a selected set of orthogonal reference states to reduce linear dependence~\cite{Stair2020}. Because HQKSD projects $\hat{H}$ itself, each Hamiltonian matrix element requires term-by-term Pauli sampling, as in VQE. Later work has sought to reduce this overhead with more efficient measurement schemes~\cite{Cohn2021,Zhang2024,LeeChoi2025}, but the underlying dependence on the Pauli decomposition of $\hat{H}$ remains. This measurement problem motivated unitary formulations of QKSD. Klymko and co-workers developed a unitary formulation of variational quantum phase estimation~\cite{Klymko2022}, and Cortes and Gray introduced the closely related unitary Krylov diagonalization method, KDM-U~\cite{Cortes2022}. Instead of projecting the Hamiltonian, these methods project the propagator $e^{-i\hat{H}\tau}$ (with $\hbar=1$). The projected propagator matrix $\mathbf{F}$ and the overlap matrix $\mathbf{S}$ can then be assembled from the same autocorrelation sequence, with only one additional time point needed for $\mathbf{F}$, so unitary QKSD (UQKSD) needs only sampled estimates of these autocorrelation values, with no term-by-term Pauli sampling of the Hamiltonian.

Resolving two energies separated by $\Delta E$ requires a total evolution time $T_{\mathrm{basis}}$ of order $1/\Delta E$, as for any sampled time series~\cite{Klymko2022}. The consequence is that closely spaced or near-degenerate states, which are important in strongly correlated molecules, require a large $T_{\mathrm{basis}}$~\cite{Mirzakhani2026}. With $M$ Krylov vectors spaced by $\tau$, the basis spans $T_{\mathrm{basis}}=(M-1)\tau$, so at fixed $T_{\mathrm{basis}}$ this resolution can be reached with either a few widely spaced vectors (large $\tau$, small $M$) or many closely spaced ones (small $\tau$, large $M$). At sufficiently small $\tau$, consecutive Krylov vectors become indistinguishable at the available noise resolution and the overlap matrix becomes increasingly ill-conditioned~\cite{Klymko2022,Shen2023,Asthana2025,Stair2020}. The resulting GEVP is therefore highly sensitive to perturbations, and the sampling-noise perturbation of the $M\times M$ overlap matrix grows with $M$, even at fixed statistical precision per measured autocorrelation~\cite{Lee2024}.

The standard remedy is thresholding~\cite{Parrish2019,Klymko2022}, which discards the eigenvectors of $\mathbf S$ whose eigenvalues fall below a cutoff and solves the GEVP in the span of the remaining ones, and perturbation bounds are available for this procedure~\cite{Epperly2022,Lee2024,Kirby2024}. A larger cutoff lowers the condition number of $\mathbf S$ restricted to the retained eigenvectors and thereby limits the amplification of noise in the GEVP. A lower condition number does not necessarily reduce the energy error, however, because a larger cutoff can also remove variational degrees of freedom needed to represent the target state~\cite{LeeChoi2025,Oliveira2026,Mirzakhani2026}. In a model calculation at fixed $T_{\mathrm{basis}}$ (Sec.~\ref{sec:illcond}), decreasing $\tau$ makes consecutive Krylov vectors more nearly linearly dependent, so more eigenvalues of $\mathbf S$ fall below the norm of the noise perturbation, $\lVert\Delta\mathbf S\rVert_2$. The smallest error obtained over a scan of thresholds then stops decreasing at the value of $M$ where the number of eigenvalues of $\mathbf S$ above $\lVert\Delta\mathbf S\rVert_2$ stops increasing. At larger $M$, cutoffs below $\lVert\Delta\mathbf S\rVert_2$ select spurious roots below the ground-state energy, and a small change of the cutoff, which does not affect the error at small $M$, can increase the error by orders of magnitude.

Keeping $M$ modest and increasing $\tau$ improves the conditioning of $\mathbf S$ and reduces the effect of noise~\cite{Klymko2022,Cortes2022}. However, because the unitary eigenvalues are $\lambda_i=e^{-iE_i\tau}$, the measured phase fixes $E_i$ only modulo $2\pi/\tau$, so standard UQKSD returns a wrapped principal-branch value once an energy lies outside $[-\pi/\tau,\pi/\tau)$. Shifting the Hamiltonian by a reliable classical estimate (e.g., the Hartree--Fock energy $E_{\mathrm{HF}}$ for weakly correlated systems) keeps the ground-state phase in the principal branch up to a larger $\tau$~\cite{Cortes2022}, but a higher state can still wrap below it and be selected as the apparent ground state (Sec.~\ref{sec:results_shift}). Conventional UQKSD therefore keeps $\tau$ small enough to remain in an unaliased region~\cite{Klymko2022,Oliveira2026}. Cortes and Gray noted that a nonzero integer multiple of $2\pi/\tau$ may have to be added to obtain energies in the desired physical range, without specifying how the integer is determined. However, no aliasing actually occurred in their own calculations~\cite{Cortes2022}.

The question of whether or not one can find a suitable $\tau$ (i.e., one that is large enough to remain decently numerically conditioned, while also small enough to not suffer from energy aliasing) becomes more difficult as one considers increasing the system size. Since the energy scale generally grows extensively with system size, even after practical reference shifts such as a Hartree–Fock shift, the $\tau$ needed to avoid aliasing will tend to decrease, thereby inevitably creating ill-conditioning problems.

\begin{figure*}[!t]
\centering
\includegraphics[width=\textwidth]{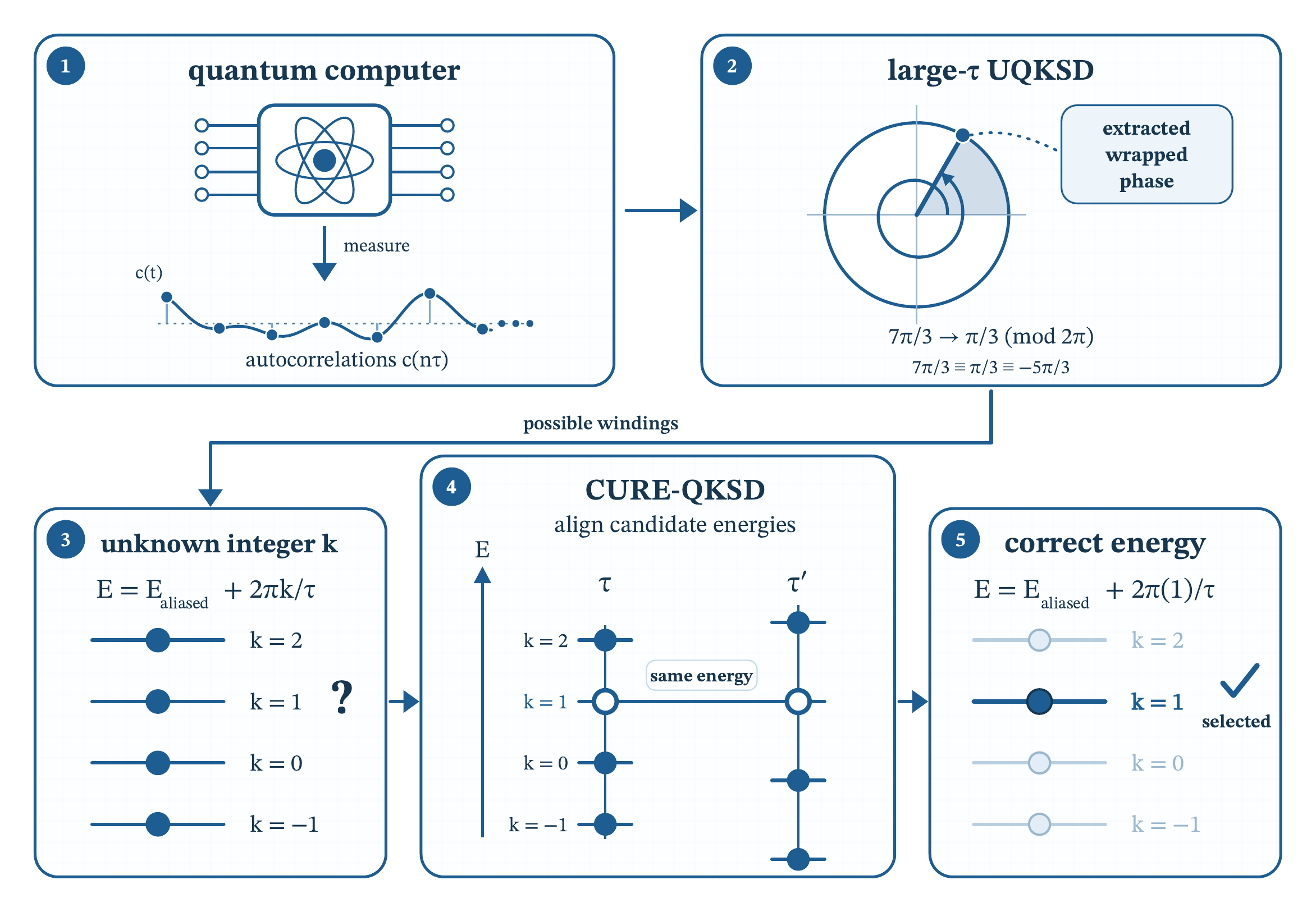}
\caption{Overview of CURE-QKSD. The UQKSD phase measured at time step $\tau$ determines the energy only up to an integer multiple $k$ of $2\pi/\tau$, so it corresponds to a set of candidate energies spaced by $2\pi/\tau$. A second time step $\tau'$ gives candidates spaced by $2\pi/\tau'$. Among the candidates shown, only the physical energy appears in both sets. This fixes $k$ for the $\tau$ run (here $k=1$) and hence the unwrapped energy.}
\label{fig:cure_overview}
\end{figure*}

The goal of this work is to develop a robust treatment of energy aliasing, so that UQKSD can be performed at large time steps that don't suffer from numerical ill-conditioning. To this end, we present Classical Unwrapping for Resolving Eigenvalues (CURE-QKSD), summarized in Fig.~\ref{fig:cure_overview}. We choose $\tau$ large enough that the target energies lie outside the principal branch, so that the required $T_{\mathrm{basis}}=(M-1)\tau$ is reached with a small $M$. CURE-QKSD leaves each UQKSD calculation and GEVP unchanged and applies the winding correction entirely as a classical post-processing step to the returned eigenphases. It first constructs a phase-consistency landscape from one UQKSD phase set over an energy window computed in advance from classical estimates (e.g., CCSD and Hartree--Fock energies). Cortes and Gray also noted that the true energies remain invariant when $\tau$ is varied whereas their aliased images move~\cite{Cortes2022}. When the landscape from one time step does not determine the winding integers uniquely, CURE-QKSD uses this property by adding a second UQKSD run at an incommensurate time step. The winding integer determined in this way is applied to the phase from the first UQKSD run. This adds an exact multiple of $2\pi/\tau$, so, provided the integer is correct, each reported energy has the same error as the energy obtained from that phase. The landscape and the second run therefore need only be accurate enough to select the integer. 

Phase-estimation algorithms resolve the same $2\pi$ ambiguity with a sequence of increasing evolution times, starting from an unaliased one, in which the estimate from each shorter time selects the branch at the next~\cite{Kitaev1995,Kimmel2015,Ding2023a}. These algorithms estimate one phase, usually that of the ground state. Dutkiewicz, Terhal, and O'Brien extended this approach to several eigenvalues: at each evolution time they estimate all phases with a time-series or matrix-pencil method, fix each winding integer by matching the estimate to one from the previous time, and choose the ratio between successive times, not necessarily an integer, so that only one image of each new estimate lies near an earlier one~\cite{Dutkiewicz2022}. Li, Ni, and Ying choose these ratios in a similar way~\cite{Li2023}. MM-QCELS fits several eigenvalues at once to the autocorrelation function sampled at random times rather than at a fixed time step, so its fit has no branch ambiguity~\cite{Ding2023b}. UQKSD, however, needs a fixed time step to build $\mathbf S$ and $\mathbf F$. CURE-QKSD also assigns winding integers to several roots, but it takes the coarse information from a classical energy window instead of shorter runs. The window fixes the winding integers; the optional second run, at a time step comparable to the first, is used only to remove phantom peaks, which are periodic images of energies that land inside the window. Every UQKSD run is therefore performed at a large $\tau$, and in the calculations reported here at most one second run was needed. CURE-QKSD has no proven error bound of this kind, and a direct comparison of its cost with these algorithms at equal quantum resources is deferred for future work. Further, because CURE-QKSD relies on extensive classical energy estimates, we expect that the number of CURE runs will increase somewhat with system size.

We first quantify the ill-conditioning produced by small $\tau$ and the resulting limits of thresholding, and then formulate the CURE-QKSD unwrapping procedure and test it on the ground and low-lying excited states of six molecular Hamiltonians. Across noiseless and noisy statevector simulations, we find that larger time steps reach a given accuracy at smaller $M$, and that CURE-QKSD correctly identifies the winding branch for states whose energies lie outside the principal branch.

\section{Theory}
\label{sec:theory}
Starting from a computationally accessible reference state $\ket{\phi_0}$ (for a molecular system, typically the Hartree--Fock determinant), QKSD builds a Krylov basis by real-time evolution under the system Hamiltonian $\hat H$,
\begin{equation}
\ket{\phi_n} = e^{-in\hat H\tau}\ket{\phi_0}, \qquad n=0,1,\dots,M-1,
\label{eq:krylov_basis}
\end{equation}
so that $\mathcal K_M = \mathrm{span}\{\ket{\phi_0},\dots,\ket{\phi_{M-1}}\}$ is spanned by $M$ equally spaced snapshots of a single trajectory. We set $\hbar=1$ throughout. For the molecular Hamiltonians we use atomic units, so energies are in Hartree, and we take chemical accuracy to be $1~\mathrm{kcal/mol}\approx1.6\times10^{-3}$~Ha. Because $\hat H$ is Hermitian, for an eigenpair $\hat H\ket{\Psi}=E\ket{\Psi}$, any function $f(\hat H)$ defined on the spectrum of $\hat H$ gives $f(\hat H)\ket{\Psi}=\lambda\ket{\Psi}$ with $\lambda=f(E)$. Expanding the sought eigenstate as $\ket{\Psi}=\sum_{n=0}^{M-1}c_n\ket{\phi_n}$ and enforcing a Galerkin (Rayleigh-Ritz) projection onto $\mathcal K_M$, i.e., requiring $\bra{\phi_j}\big(f(\hat H)-\lambda\big)\ket{\Psi}=0$ for every basis state, gives one generalized eigenvalue problem (GEVP) for any choice of $f$:
\begin{equation}
\begin{split}
\mathbf F\mathbf c &= \lambda\, \mathbf S\mathbf c, \\
F_{jn}&=\bra{\phi_j}f(\hat H)\ket{\phi_n}, \\
S_{jn}&=\braket{\phi_j}{\phi_n}.
\end{split}
\label{eq:gevp}
\end{equation}
The overlap matrix does not depend on $f$. Using Eq.~\eqref{eq:krylov_basis},
\begin{equation}
\begin{split}
S_{jn}&=\bra{\phi_0}e^{ij\hat H\tau}e^{-in\hat H\tau}\ket{\phi_0} \\
&= \bra{\phi_0}e^{-i(n-j)\hat H\tau}\ket{\phi_0} \equiv c(n-j),
\end{split}
\label{eq:S_toeplitz}
\end{equation}
so $\mathbf S$ is a Toeplitz matrix built entirely from the autocorrelation function $c(k)=\bra{\phi_0}e^{-ik\hat H\tau}\ket{\phi_0}$, which is measurable one integer $k$ at a time with a controlled-evolution (Hadamard-test-type) circuit~\cite{Kitaev1995}. The choice of $f$ changes only $\mathbf F$, and with it the measurement cost and the meaning of $\lambda$.

\paragraph{Hamiltonian QKSD.} Taking $f(\hat H)=\hat H$ gives $\lambda=E$ directly, with
\begin{equation}
F_{jn}=\bra{\phi_0}e^{ij\hat H\tau}\,\hat H\,e^{-in\hat H\tau}\ket{\phi_0}.
\end{equation}
Because of the extra factor of $\hat H=\sum_l h_l\hat P_l$ inserted between the two evolutions, $F_{jn}$ cannot be written in terms of $c(k)$ alone: each distinct value $\bra{\phi_0}\hat H\,e^{-ik\hat H\tau}\ket{\phi_0}$ requires its own set of Pauli measurements, reproducing the sampling overhead discussed in Sec.~\ref{sec:intro}.

\paragraph{Unitary QKSD.} Taking instead $f(\hat H)=e^{-i\hat H\tau}$, the propagator itself, gives $\lambda_i=e^{-iE_i\tau}$ and
\begin{equation}
\begin{split}
F_{jn}&=\bra{\phi_0}e^{ij\hat H\tau}e^{-i\hat H\tau}e^{-in\hat H\tau}\ket{\phi_0}\\
&= c(n+1-j),
\end{split}
\label{eq:F_from_S}
\end{equation}
Thus $\mathbf F$ is the one-step-shifted Toeplitz array generated by the same autocorrelation data as $\mathbf S$, with the extra value $c(M)$ needed at the boundary. Both matrices are constructed from $\{c(0),c(1),\dots,c(M)\}$, which is why UQKSD needs no separate measurement of $\hat H$: this is the origin of the measurement saving quoted in Sec.~\ref{sec:intro}. 

Klymko and co-workers use this shared structure, recovering the propagator matrix from the autocorrelation data rather than measuring it independently~\cite{Klymko2022}. In signal processing, this kind of eigenvalue problem, built from two matrices of samples shifted by one time step, is known as the matrix-pencil method and is used to extract the complex exponentials in a sampled signal~\cite{Hua1990}. O'Brien, Tarasinski, and Terhal used it to estimate several eigenphases at once from single-ancilla measurements~\cite{OBrien2019}. Cortes and Gray compare HQKSD (their KDM-H) with UQKSD (their KDM-U) and show that the number of quantum-computer calls required by HQKSD scales with $L$, the number of Pauli terms in $\hat H$, whereas the number required by UQKSD does not depend on $L$~\cite{Cortes2022}. They also computed the energy of the fourth singlet excited state of H$_2$O at $20$ O--H bond lengths with ten time steps, and UQKSD reached chemical accuracy at $9$ of these bond lengths and HQKSD at $4$~\cite{Cortes2022}.

Since $\lambda_i$ ideally lies on the unit circle, the energy is recovered from its phase,
\begin{equation}
\lambda_i = e^{-iE_i\tau} \;\;\Longrightarrow\;\; E_i = -\frac{1}{\tau}\arg(\lambda_i),
\label{eq:phase_extraction}
\end{equation}
where $\arg(\cdot)\in(-\pi,\pi]$ denotes the principal branch. The reported energy therefore lies in $[-\pi/\tau,\pi/\tau)$ and determines the physical energy only modulo $2\pi/\tau$; an energy outside this interval is returned as its image $E_i-2\pi k/\tau$, with integer $k$, inside the interval, a phenomenon we call \emph{phase aliasing}. For an unshifted Hamiltonian, the ground state first becomes aliased once
\begin{equation}
\tau > \tau_{\mathrm{alias}} = \frac{\pi}{|E_{\mathrm{gs}}|}.
\label{eq:tau_alias}
\end{equation}
Throughout, phase aliasing means only the branch ambiguity in the eigenvalues returned by the GEVP. The separate case in which two eigenstates acquire the same phase factor in every Krylov vector is a property of the basis, which CURE-QKSD does not change (Appendix~\ref{app:basis_alias}).

\subsection{Ill-conditioning and the limits of thresholding}
\label{sec:illcond}
As noted in Sec.~\ref{sec:intro}, resolving eigenvalues separated by $\Delta E$ requires $T_{\mathrm{basis}}=(M-1)\tau$ of order $1/\Delta E$~\cite{Mirzakhani2026}. At fixed $T_{\mathrm{basis}}$, this resolution can be reached either with a few widely spaced Krylov vectors (large $\tau$, small $M$) or with many closely spaced ones (small $\tau$, large $M$). Constructing $\mathbf F$ additionally requires $c(M)$ at $T_{\max}=M\tau$, but that extra point does not extend the $M$-vector basis itself, whose total evolution time remains $T_{\mathrm{basis}}$. We now show quantitatively why the second option becomes unreliable in the presence of noise.

For a normalized reference state, the overlap of consecutive Krylov vectors is $\langle\phi_n|\phi_{n+1}\rangle=c(1)$, and expanding $c(1)$ in powers of $\tau$ gives
\begin{equation}
|\langle\phi_n|\phi_{n+1}\rangle|^2=1-\sigma_E^2\tau^2+O(\tau^4),
\label{eq:consecutive_overlap}
\end{equation}
where $\sigma_E^2=\bra{\phi_0}\hat H^2\ket{\phi_0}-\bra{\phi_0}\hat H\ket{\phi_0}^2$ is the energy variance of the reference state. Consecutive Krylov vectors are therefore nearly parallel when $\sigma_E\tau\ll1$. To see how this affects the conditioning of the overlap matrix, consider the $2\times2$ principal submatrix formed by two consecutive Krylov vectors,
\begin{equation}
\mathbf S^{(2)}=
\begin{pmatrix}
1 & c(1)\\
c(1)^* & 1
\end{pmatrix}.
\end{equation}
Its eigenvalues are $\lambda_\pm=1\pm|c(1)|$, giving the spectral condition number
\begin{equation}
\kappa_2(\mathbf S^{(2)})
=\frac{1+|c(1)|}{1-|c(1)|}
=\frac{4}{\sigma_E^2\tau^2}+O(1),
\end{equation}
for nonzero $\sigma_E$ in the small-$\tau$ limit. Since $\mathbf S^{(2)}$ is a principal submatrix of the full overlap matrix $\mathbf S$, eigenvalue interlacing gives $\lambda_{\max}(\mathbf S)\geq\lambda_+$ and $\lambda_{\min}(\mathbf S)\leq\lambda_-$. Thus, for nonsingular $\mathbf S$,
\begin{equation}
\kappa_2(\mathbf S)\geq\kappa_2(\mathbf S^{(2)})
=\frac{4}{\sigma_E^2\tau^2}+O(1).
\end{equation}
This bound shows that small $\sigma_E\tau$ necessarily leads to poor conditioning of $\mathbf S$, quantifying the effect of near-linear dependence among consecutive Krylov vectors. Near-linear dependence has also been identified as a source of overlap-matrix ill-conditioning in previous studies~\cite{Seki2021,Shen2023,Asthana2025}.

Lee, Lee, and Huh showed that the Krylov basis becomes increasingly linearly dependent when $\tau\ll\lVert\hat H\rVert_2^{-1}$, where $\lVert\cdot\rVert_2$ denotes the spectral norm~\cite{Lee2024}. This condition can be expressed in terms of the spectral width because a constant shift of $\hat H$ changes each Krylov vector only by a phase and leaves the eigenvalues of $\mathbf S$ unchanged. The least restrictive bound is obtained by shifting $\hat H$ by $\alpha=(E_{\max}+E_{\min})/2$, giving $\tau(E_{\max}-E_{\min})/2\ll1$, where $E_{\max}$ and $E_{\min}$ are the extremal eigenvalues of $\hat H$. Since $\sigma_E\leq(E_{\max}-E_{\min})/2$ for any normalized reference state, this condition also implies $\sigma_E\tau\ll1$.

In exact arithmetic, nearly parallel Krylov vectors are generically still linearly independent, but numerical and measurement noise limits how small a difference between them can be resolved. Klymko and co-workers define a resolution $\eta_{\mathrm{noise}}$ such that two measured or computed values are distinguishable only when they differ by more than $\eta_{\mathrm{noise}}$~\cite{Klymko2022}. If $\Delta E_{\min}$ is the smallest energy gap between eigenstates in the support of $\ket{\phi_0}$, one time step adds a relative phase $\tau\Delta E_{\min}$ between the two eigenstates separated by this gap. Klymko and co-workers estimate that a new time-evolved vector does not improve the energy estimates when this phase is below the resolution, so finite noise sets a minimum useful time step, $\tau>\eta_{\mathrm{noise}}/\Delta E_{\min}.$ In the numerical tests of Stair, Huang, and Evangelista, the condition number of a single-reference overlap matrix at fixed $\tau$ grows from $3\times10^{5}$ at $M=4$ to $4\times10^{17}$ at $M=20$, which introduces numerical instability into the GEVP that can be amplified further by gate infidelity and measurement error~\cite{Stair2020}. Epperly, Lin, and Nakatsukasa, citing these same observations, note that small time steps worsen the ill-conditioning of the projected Hamiltonian and overlap matrices, while quantum-subspace methods of this kind often perform better with larger ones~\cite{Epperly2022}.

\begin{figure*}
\centering
\includegraphics[width=\textwidth]{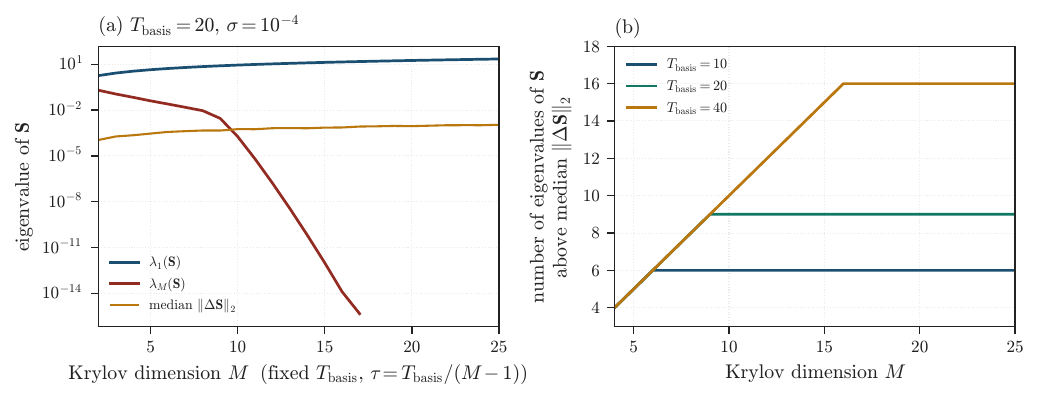}
\caption{Eigenvalues of $\mathbf S$ for the 48-level toy model with autocorrelation noise $\sigma=10^{-4}$. (a) Largest and smallest eigenvalues of the exact $\mathbf S$ versus $M$ at fixed $T_{\mathrm{basis}}=(M-1)\tau=20$, compared with the median of $\lVert\Delta\mathbf S\rVert_2$ over 31 noise realizations (gold). The smallest eigenvalue falls below the median $\lVert\Delta\mathbf S\rVert_2$ from $M=10$. (b) Number of eigenvalues of the exact $\mathbf S$ larger than the median $\lVert\Delta\mathbf S\rVert_2$ versus $M$. This number becomes constant at $6$, $9$, and $16$ for $T_{\mathrm{basis}}=10$, $20$, and $40$, respectively.}
\label{fig:conditioning}
\end{figure*}

\begin{figure*}
\centering
\includegraphics[width=\textwidth]{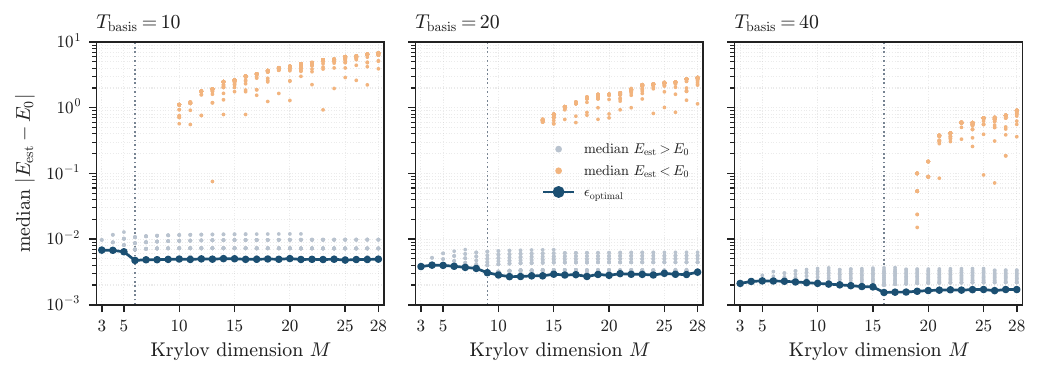}
\caption{Median ground-state error (31 noise realizations) versus $M$ for $T_{\mathrm{basis}}=10$, $20$, and $40$ (left to right), for the same spectrum and autocorrelation-noise model as Fig.~\ref{fig:conditioning}. Points show the median error at each of the 51 scanned values of $\epsilon_{\mathrm{thresh}}$ (logarithmically spaced from $10^{-6}$ to $10^{-1}$). They are gray when the median estimate lies above $E_0$ and orange when it lies below $E_0$, where no level exists. The line connects the smallest of them, at $\epsilon_{\mathrm{optimal}}$. The line flattens at the dotted vertical lines, $M=6$, $9$, and $16$, where the number of eigenvalues of $\mathbf S$ above the median $\lVert\Delta\mathbf S\rVert_2$ stops increasing [Fig.~\ref{fig:conditioning}(b)]. Orange points appear from $M=10$, $14$, and $19$, all at cutoffs below the median $\lVert\Delta\mathbf S\rVert_2$.}
\label{fig:threshold_trap}
\end{figure*}

\begin{figure*}
\centering
\includegraphics[width=\textwidth]{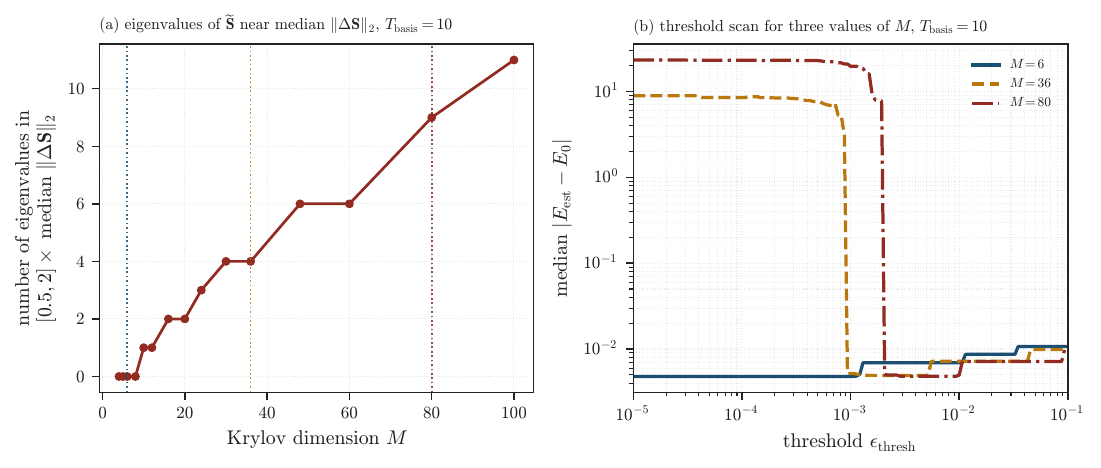}
\caption{Thresholding at fixed $T_{\mathrm{basis}}=(M-1)\tau=10$ with autocorrelation noise $\sigma=10^{-4}$. (a) Median number of eigenvalues of $\widetilde{\mathbf S}$ between $0.5$ and $2$ times the median $\lVert\Delta\mathbf S\rVert_2$ versus $M$ (31 noise realizations). Dotted lines mark the values of $M$ used in (b). (b) Median ground-state error (21 noise realizations) versus $\epsilon_{\mathrm{thresh}}$ for $M=6$, $36$, and $80$ ($\tau=2.000$, $0.286$, and $0.127$), for which the median $\lVert\Delta\mathbf S\rVert_2$ is $3.70\times10^{-4}$, $1.29\times10^{-3}$, and $2.21\times10^{-3}$. At $M=6$ the error changes by less than a factor of $2.3$ over the scan, whereas at $M=36$ and $80$ it increases by more than two orders of magnitude when $\epsilon_{\mathrm{thresh}}$ falls below $0.73$ and $0.93$ times the median $\lVert\Delta\mathbf S\rVert_2$, where a spurious root is selected.}
\label{fig:epsilon_sweep}
\end{figure*}

To show how the eigenvalues of $\mathbf S$ and the thresholded error change with $\tau$ at fixed $T_{\mathrm{basis}}$, Figs.~\ref{fig:conditioning}--\ref{fig:epsilon_sweep} use one toy model: $N=48$ levels $E_\ell=-1+2[\ell/(N-1)]^2$, $\ell=0,\ldots,N-1$, and the reference state $\ket{\phi_0}=\sum_\ell\sqrt{w_\ell}\ket{E_\ell}$. The normalized weights satisfy $w_\ell\propto b_\ell\exp[-(E_\ell-E_0)/0.5]$ with $b_\ell=1+N/(\ell+1)^2$, producing a dominant low-energy component and a decaying tail. The spectrum and weights are chosen by hand and are not taken from a molecule. The reference state has a weight of about $0.50$ on the ground state, about $0.33$ in total on the next seven states, and the remaining weight on the other 40 levels. Figure~\ref{fig:conditioning}(a) shows that, at fixed $T_{\mathrm{basis}}=(M-1)\tau=20$, increasing $M$ from $7$ to $17$ increases the largest eigenvalue of $\mathbf S$ from $6.17$ to $15.1$, while the smallest decreases from $1.51\times10^{-2}$ to $4.58\times10^{-16}$, so the exact condition number grows from $4.08\times10^{2}$ to $3.29\times10^{16}$. Thus only ten additional, more closely spaced vectors worsen the conditioning by nearly fourteen orders of magnitude. Because every dimension in Fig.~\ref{fig:conditioning} satisfies $M\leq25<48$, where 48 is the number of levels with nonzero weight in $\ket{\phi_0}$, the decrease of the smallest eigenvalue is not a consequence of using more Krylov vectors than populated levels but of the near-linear dependence of consecutive Krylov vectors at smaller $\tau$.

This ill-conditioning alone is a reason to keep $M$ modest. In addition, the noise entering $\mathbf S$ does not stay fixed as $M$ grows. Because $\mathbf S$ is Toeplitz [Eq.~\eqref{eq:S_toeplitz}], we add noise to the autocorrelation values rather than to individual matrix entries: each $c(k)$ with $k\neq0$ receives an independent complex Gaussian error of standard deviation $\sigma$, and every entry $S_{jn}$ with $|n-j|=k$ contains that error or its complex conjugate. Thus $\Delta\mathbf S$ remains Hermitian and Toeplitz, built from $M-1$ random lag errors rather than $M^2$ independent entries. A tilde denotes a matrix built from noisy autocorrelation values, such as the noisy overlap matrix $\widetilde{\mathbf S}=\mathbf S+\Delta\mathbf S$. Both $\mathbf S$ and $\widetilde{\mathbf S}$ are Hermitian, and Weyl's inequality (Appendix~\ref{app:weyl}) bounds the change in each eigenvalue by the spectral norm of the perturbation, with the eigenvalues of each matrix in decreasing order,
\begin{equation}
\left|\lambda_j(\widetilde{\mathbf S})-\lambda_j(\mathbf S)\right|
\leq \lVert\Delta\mathbf S\rVert_2,
\qquad j=1,\ldots,M.
\label{eq:weyl_overlap}
\end{equation}

The correlations between repeated Toeplitz entries affect the size and distribution of $\lVert\Delta\mathbf S\rVert_2$, but not Eq.~\eqref{eq:weyl_overlap}. 
Lee, Lee, and Huh derive non-asymptotic 
bounds on this sampling-induced perturbation directly from the measurement statistics and find that the bound grows with the Krylov dimension~\cite{Lee2024}. Because the exact $c(k)$ cancel in $\Delta\mathbf S=\widetilde{\mathbf S}-\mathbf S$, $\Delta\mathbf S$ is built only from the $M-1$ added errors, so $\lVert\Delta\mathbf S\rVert_2$ depends on $\sigma$ and $M$ but not on the spectrum or $\tau$. Figure~\ref{fig:conditioning}(a) shows the median of $\lVert\Delta\mathbf S\rVert_2$ over 31 noise realizations.
Figure~\ref{fig:conditioning}(b) shows the number of eigenvalues of the exact $\mathbf S$ that are larger than the median $\lVert\Delta\mathbf S\rVert_2$, as a function of $M$ at fixed $T_{\mathrm{basis}}$. This number stops increasing at $6$, $9$, and $16$ for $T_{\mathrm{basis}}=10$, $20$, and $40$, respectively. 
To summarize, increasing the number of Krylov vectors ($M$) while keeping total evolution time constant increases only the dimension of the GEVP but not the number of eigenvalues of the exact $\mathbf S$ larger than the median $\lVert\Delta\mathbf S\rVert_2$, meaning that the additional Krylov vectors only add directions that are indistinguishable from noise, and that the accuracy is limited only by evolution time, $T_\text{basis}$.

The thresholding method, also called canonical orthogonalization~\cite{Lowdin1967}, diagonalizes $\widetilde{\mathbf S}$, discards the eigenvectors whose eigenvalues fall below a cutoff $\epsilon_{\mathrm{thresh}}$, and solves the GEVP in the retained subspace. This is useful but introduces a trade-off between sensitivity to noise and truncation error. A small $\epsilon_{\mathrm{thresh}}$ retains eigenvectors of $\widetilde{\mathbf S}$ whose eigenvalues are comparable to or smaller than $\lVert\Delta\mathbf S\rVert_2$ and are therefore dominated by noise~\cite{Klymko2022,Epperly2022,Lee2024,Kirby2024}, whereas a large $\epsilon_{\mathrm{thresh}}$ is overly conservative and removes physically meaningful degrees of freedom that are needed to represent the target eigenstate~\cite{LeeChoi2025,Mirzakhani2026,Oliveira2026,DeVriendt2026}. Kirby's rigorous HQKSD bounds highlight the difficulty in choosing a suitable threshold value, $\epsilon_{\mathrm{thresh}}$~\cite{Kirby2024}. Subject to the stability requirement $\epsilon_{\mathrm{thresh}}\geq\lVert\Delta\mathbf S\rVert_2$, the positive-error upper bound is tightest at the smallest admissible cutoff and loosens as $\epsilon_{\mathrm{thresh}}$ grows, because a larger part of the Krylov subspace is discarded. The corresponding lower bound contains a term proportional to $-1/\epsilon_{\mathrm{thresh}}$, so it is weakest for $\epsilon_{\mathrm{thresh}}\approx\lVert\Delta\mathbf S\rVert_2$ and becomes tighter as $\epsilon_{\mathrm{thresh}}$ increases. The smallest admissible cutoff therefore gives the tightest upper bound but the weakest lower bound on the energy error. Although those bounds apply to HQKSD rather than the non-Hermitian UQKSD GEVP, they show that raising the threshold reduces noise amplification but increases the truncation error.

That still leaves the question the next two figures answer: at fixed evolution time ($T_{\mathrm{basis}}$), increasing $M$ decreases $\tau=T_{\mathrm{basis}}/(M-1)$. \textit{Does thresholding let us keep reducing $\tau$ in this way and still lower the ground-state error, or does the error eventually stop decreasing, or even grow?} At each $M$ in Fig.~\ref{fig:threshold_trap}, we scan $\epsilon_{\mathrm{thresh}}$ and denote by $\epsilon_{\mathrm{optimal}}$ the value that gives the smallest median error. The error at $\epsilon_{\mathrm{optimal}}$ is the best result thresholding can give at that $M$, so if it stops decreasing as $M$ grows, no choice of cutoff can make the extra vectors useful. Note that this is purely a diagnostic test and not a practical approach since finding $\epsilon_\text{optimal}$ requires the exact $E_0$. All three panels use $M=3,\ldots,28$, with extra care taken for the small $M$ values \footnote{At the small-$M$ points for which $\tau>\pi$ and the bounded test spectrum no longer fits in one principal branch, we apply the CURE correction of Sec.~\ref{sec:cure_methodology}, with a second run at $\tau/\varphi$, before evaluating the physical energy error. This changes only the winding assigned to the final UQKSD eigenvalue, not either thresholded GEVP. The figure therefore compares thresholding over the same $M$ domain in all panels without confusing a branch-wrapping error with a thresholding error.}.

In Fig.~\ref{fig:threshold_trap}, the median error at $\epsilon_{\mathrm{optimal}}$ falls by more than $10\%$ for the last time at $M=6$, $9$, and $16$ for $T_{\mathrm{basis}}=10$, $20$, and $40$, respectively, and additional vectors through $M=28$ produce no sustained improvement. Between $M=3$ and $28$, this error decreases by only a factor of $1.4$, $1.2$, and $1.2$. The plateaus begin at the same values of $M$ at which the count in Fig.~\ref{fig:conditioning}(b) stops increasing (dotted lines), meaning that the accuracy is ultimately determined by the number of overlap matrix eigenvalues that are greater than the noise, which we just observed reaches a max value determined by the noise. Once that count stops increasing, the accuracy no longer improves, so additional vectors beyond that point, and the corresponding further reductions of $\tau$ at fixed $T_{\mathrm{basis}}$, give no further improvement. 

At every $M$, the scanned thresholds whose median estimate lies above $E_0$ (gray points) give errors within a factor of $2.6$ of $\epsilon_{\mathrm{optimal}}$. Every larger error comes from a spurious root below $E_0$ (orange points) and is $9$ to $1400$ times the error at $\epsilon_{\mathrm{optimal}}$. In single realizations, spurious roots first appear at $M=7$, $11$, and $17$, just after the dotted lines, and no scanned cutoff gives an estimate below $E_0$ at smaller $M$. Orange points occur only at cutoffs below the median $\lVert\Delta\mathbf S\rVert_2$, and at $M=28$ they occur at $29$ of the $31$ such cutoffs in each panel. Even at $\epsilon_{\mathrm{optimal}}$, a spurious root is returned in up to $13$ of the $31$ realizations at large $M$ ($T_{\mathrm{basis}}=20$, $M=20$).

Figure~\ref{fig:epsilon_sweep} shows that adding vectors beyond the plateau does not lower the best error but makes the result depend sharply on the cutoff, because more eigenvalues of $\widetilde{\mathbf S}$ lie near $\lVert\Delta\mathbf S\rVert_2$. At $T_{\mathrm{basis}}=10$, it compares $M=6$, $36$, and $80$ ($\tau=2.000$, $0.286$, and $0.127$), where $M=6$ is the start of the plateau in Fig.~\ref{fig:threshold_trap}. Panel (a) shows the number of eigenvalues of $\widetilde{\mathbf S}$ between $0.5$ and $2$ times the median $\lVert\Delta\mathbf S\rVert_2$. The median count is $0$ at $M=6$, where all six exact overlap eigenvalues exceed the median $\lVert\Delta\mathbf S\rVert_2$, but grows to $4$ at $M=36$ and $9$ at $M=80$. Lowering the cutoff from twice to half the median $\lVert\Delta\mathbf S\rVert_2$ therefore adds no eigenvectors to the retained subspace at $M=6$, but adds about $4$ and $9$ at $M=36$ and $80$. Panel (b) shows the median ground-state error as a function of $\epsilon_{\mathrm{thresh}}$ between $10^{-5}$ and $10^{-1}$. At $M=6$, the median error lies between $4.79\times10^{-3}$ and $1.07\times10^{-2}$ at every $\epsilon_{\mathrm{thresh}}$ in the scan. At $M=36$ and $80$, it is below $10^{-2}$ only for $\epsilon_{\mathrm{thresh}}\geq9.36\times10^{-4}$ and $\epsilon_{\mathrm{thresh}}\geq2.06\times10^{-3}$, respectively, which are $0.73$ and $0.93$ times the median $\lVert\Delta\mathbf S\rVert_2$ ($1.29\times10^{-3}$ and $2.21\times10^{-3}$). At the next scanned value, $6\%$ lower, the median errors are $3.09$ and $7.74$, because the ground-state estimate comes from a spurious root with an energy lower than $E_0$. Epperly, Lin, and Nakatsukasa noted that noise can easily introduce such roots when a Hermitian GEVP is nearly singular~\cite{Epperly2022}. With $\epsilon_{\mathrm{thresh}}$ equal to the median $\lVert\Delta\mathbf S\rVert_2$, a spurious root is selected in $2$ of $21$ noise realizations at $M=36$ and in $6$ of $21$ at $M=80$, whereas at $M=6$ none of the $21$ realizations selects one at half, once, or twice this value. At $\epsilon_{\mathrm{optimal}}$, the median errors at $M=36$ and $80$ are $4.92\times10^{-3}$ and $4.82\times10^{-3}$, no smaller than at $M=6$. At larger $M$, the result therefore depends sharply on where the cutoff is placed relative to $\lVert\Delta\mathbf S\rVert_2$, which varies between noise realizations and is not known exactly in practice. In Fig.~\ref{fig:threshold_trap}, spurious roots appear at the same $T_{\mathrm{basis}}$ from $M=7$, far below the 48 levels populated in $\ket{\phi_0}$, so they do not require $M$ to exceed the number of populated levels.

Lee, Lee, and Huh observed the same behavior in noisy HQKSD simulations of the one-dimensional Fermi--Hubbard model: as more eigenvectors of $\widetilde{\mathbf S}$ are retained, the ground-state error first decreases and then increases sharply~\cite{Lee2024}. At a fixed Krylov dimension, they computed the ground-state error as a function of the number of eigenvectors of $\widetilde{\mathbf S}$ retained after thresholding. This scan is equivalent to the $\epsilon_{\mathrm{thresh}}$ scan in Fig.~\ref{fig:epsilon_sweep}(b), because decreasing $\epsilon_{\mathrm{thresh}}$ increases the number of retained eigenvectors. Without noise, their error decreases as more eigenvectors are retained. With sampling noise, the error follows the noiseless result until about the point where the retained eigenvalues fall below their bound on $\lVert\Delta\mathbf S\rVert_2$. Below this level, they find that the eigenvalues of $\widetilde{\mathbf S}$ differ from those of $\mathbf S$. More shots reduce $\lVert\Delta\mathbf S\rVert_2$, so the increase starts at a larger number of retained eigenvectors. For $M=36$ and $80$ in Fig.~\ref{fig:epsilon_sweep}(b), decreasing $\epsilon_{\mathrm{thresh}}$ likewise first reduces the error and then increases it by more than two orders of magnitude, once eigenvectors whose eigenvalues are comparable to $\lVert\Delta\mathbf S\rVert_2$ are retained. In both calculations, an additional eigenvector reduces the error only if its eigenvalue is large compared with $\lVert\Delta\mathbf S\rVert_2$. An eigenvalue comparable to $\lVert\Delta\mathbf S\rVert_2$ can change by an amount comparable to itself [Eq.~\eqref{eq:weyl_overlap}], and retaining its eigenvector increases the error.

Together, Figs.~\ref{fig:conditioning}--\ref{fig:epsilon_sweep} show the effect of increasing $M$, and thus decreasing $\tau$, at fixed $T_{\mathrm{basis}}$ in this toy model. The smallest eigenvalues of $\mathbf S$ decrease, and the number of its eigenvalues larger than the median $\lVert\Delta\mathbf S\rVert_2$ stops increasing beyond a value of $M$ that increases with $T_{\mathrm{basis}}$ (Fig.~\ref{fig:conditioning}). At the same values of $M$, the error at $\epsilon_{\mathrm{optimal}}$ stops decreasing (Fig.~\ref{fig:threshold_trap}). One or two dimensions later, cutoffs below the median $\lVert\Delta\mathbf S\rVert_2$ begin to select spurious roots below $E_0$, and these account for every large error in the scan. At larger $M$, more eigenvalues of $\widetilde{\mathbf S}$ lie near $\lVert\Delta\mathbf S\rVert_2$, and a small change of the cutoff, which does not affect the error at $M=6$, can select such a root (Fig.~\ref{fig:epsilon_sweep}). Increasing $\tau$ instead makes consecutive Krylov vectors less nearly parallel [Eq.~\eqref{eq:consecutive_overlap}], and gives a smaller UQKSD problem whose result is less sensitive to the threshold. This is consistent with Oliveira, Ziems, and Glaser's UQKSD simulations, which find more accurate energies at fixed Krylov dimension for larger time steps within their unaliased regime~\cite{Oliveira2026}.

\subsection{CURE-QKSD methodology}
\label{sec:cure_methodology}

In this section, we introduce a post-processing method for UQKSD calculations that allows the use of large time steps to avoid the numerical ill-conditioning problems described in the previous section. By keeping $M$ modest and choosing $\tau>\tau_{\mathrm{alias}}$ [Eq.~\eqref{eq:tau_alias}], one avoids the poorly conditioned small-$\tau$ regime of Sec.~\ref{sec:illcond}, but it reopens the aliasing problem described in Sec.~\ref{sec:theory}: each measured phase $\theta_i\equiv\arg(\lambda_i)$ fixes $E_i$ only modulo $2\pi/\tau$. Classical Unwrapping for Resolving Eigenvalues (CURE-QKSD) resolves this ambiguity entirely in classical post-processing, without changing any UQKSD GEVP. 

CURE-QKSD consists of the following steps:
\begin{enumerate}
\item Run UQKSD at a base time step $\tau_1>\tau_{\mathrm{alias}}$ and keep the phases $\theta_i^{(1)}$ that pass a validation test.
\item Obtain an energy window $\mathcal W$ from classical estimates.
\item Build the phase-consistency landscape $L_1(E)$ over $\mathcal W$, which equals $1$ at each candidate energy $E=(-\theta_i^{(1)}+2\pi k)/\tau_1$ of a phase $\theta_i^{(1)}$.
\item Unless $\mathcal W$ is narrower than $2\pi/\tau_1$ and $L_1$ has exactly one peak per target state, run UQKSD at a second time step $\tau_2=\tau_1/\varphi$ and average the two landscapes. A true energy is a peak of both landscapes, whereas a phantom peak generally is not, so only the peaks on which both runs agree remain. Further UQKSD runs at different time steps can be added in the same way until a new run removes no peak.
\item Use the remaining peaks as guides to the target energies. For each target, unwrap the phase of a base-run root to its candidate energy nearest the guide, $E=(-\theta_i^{(1)}+2\pi k_i)/\tau_1$, with the winding number $k_i$ of Eq.~\eqref{eq:winding}.
\end{enumerate}
Only steps 1 and 4 use the quantum computer. Because each reported energy is a base-run energy shifted by an exact multiple of $2\pi/\tau_1$, it has the accuracy of the base run if $k_i$ is correct. The rest of this section describes the validation test, the window, how the landscapes are built and compared, and how the integers $k_i$ are assigned.

\paragraph{Base run and root validation.} We first solve UQKSD [Eq.~\eqref{eq:gevp} with $f=e^{-i\hat H\tau}$] at a base time step $\tau_1>\tau_{\mathrm{alias}}$, chosen large enough to mitigate the small-step conditioning problem at a modest $M$. Even at this larger step, not every eigenvalue returned by the thresholded GEVP necessarily corresponds to a physically meaningful state: numerical artifacts of the reduced problem can produce roots near the unit circle that carry negligible weight in the reference state. For a unitary QKSD projection, deviation of a recovered root from the unit circle is itself a useful diagnostic of solution reliability~\cite{Oliveira2026}. We therefore keep only roots $\lambda_i$ whose magnitude is close to 1, $|\,|\lambda_i|-1\,|<\delta$, and further validate each remaining root by its spectral weight. The ideal noiseless autocorrelation is a sum of undamped unit-modulus exponentials, $c(k)=\sum_i a_i\lambda_i^{\,k}$ with $a_i = \left|\langle E_i\mid\phi_0 \rangle\right|^2$ and $|\lambda_i|=1$. For the measured sequence we fit the same exponential form to the remaining roots by linear least squares against their Vandermonde matrix. Roots with $|a_i|<c_{\mathrm{val}}\max_j|a_j|$, where $c_{\mathrm{val}}$ is the validation cutoff, are discarded as artifacts rather than treated as physical states. The same amplitude test is used to reject spurious roots in matrix-pencil phase estimation~\cite{OBrien2019,Dutkiewicz2022}. The phases of the remaining roots are denoted $\{\theta_i^{(1)}\}$. From here on, ``root'' means a root that passes this test.

\paragraph{Classical energy window.} Each phase $\theta_i^{(1)}$ is consistent with an infinite ladder of candidate energies, $E=(-\theta_i^{(1)}+2\pi k)/\tau_1$ for integer $k$, so a finite energy window $\mathcal W$ is needed to select the branch. We define $\mathcal W$ using only classically accessible quantities. Let $E_{\mathrm{CCSD}}$ be the coupled-cluster singles-and-doubles energy~\cite{Purvis1982}, let $E_{\mathrm{HF}}=\bra{\phi_0}\hat H\ket{\phi_0}$, and let $\Delta_{\mathrm{HL}}$ be the gap between the canonical Hartree--Fock LUMO and HOMO orbital energies. 
Because CCSD is not variational and can lie above the exact ground-state energy, we place the lower edge a buffer $\delta_W$ below $E_{\mathrm{CCSD}}$. The ground-state calculations use
\begin{equation}
\mathcal W_{\mathrm{gs}}=[E_{\mathrm{CCSD}}-\delta_W,\ E_{\mathrm{HF}}],
\label{eq:window_gs}
\end{equation}
where the upper edge is the variational Hartree--Fock bound. For calculations targeting several low-lying states, we raise the upper edge by $\Delta_{\mathrm{HL}}+\delta_W$, using the HOMO--LUMO gap as a simple estimate of the excitation energy. CIS~\cite{Foresman1992} or EOM-CCSD~\cite{Stanton1993} excitation energies may give better excited-state edges, but we use this simpler estimate here:
\begin{equation}
\mathcal W_{\mathrm{multi}}=[E_{\mathrm{CCSD}}-\delta_W,\ E_{\mathrm{HF}}+\Delta_{\mathrm{HL}}+\delta_W].
\label{eq:window_multi}
\end{equation}
The window is an assumption of the method. The upper edge of $\mathcal W_{\mathrm{gs}}$ is the variational Hartree--Fock bound, but the lower edge holds only if $E_{\mathrm{CCSD}}-\delta_W$ lies below the exact ground-state energy, which can fail for strongly correlated systems. In this study, we use the same fixed $\delta_W$ for every system (Sec.~\ref{sec:methodology}). However, a fixed buffer is not size-extensive, so for larger systems a size-extensive buffer, such as a fixed fraction of the CCSD correlation energy $E_{\mathrm{HF}}-E_{\mathrm{CCSD}}$, could be used instead. A window with rigorous ground-state edges could also be built from a truncated CI energy, such as CISD, and a variational two-electron reduced-density-matrix (v2RDM) energy, which bound the exact energy from above and below, respectively~\cite{Mazziotti2012}. The upper edge of $\mathcal W_{\mathrm{multi}}$ is not a bound: $\Delta_{\mathrm{HL}}$ refers only to the HOMO--LUMO excitation, and a targeted state can be an excitation to a higher orbital or a double excitation. A target energy outside $\mathcal W$ cannot be recovered.

\begin{figure}
\centering
\includegraphics[width=\columnwidth]{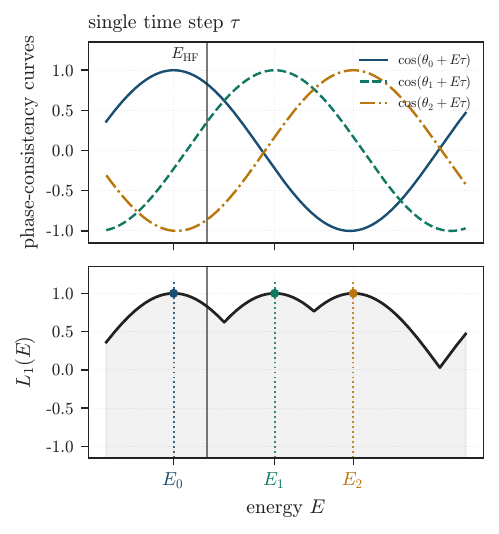}
\caption{Ideal noiseless phase-consistency landscape [Eq.~\eqref{eq:landscape}] for three toy energy levels and a single time step $\tau$. Top: the three per-root phase-consistency curves $\cos(\theta_i+E\tau)$. Bottom: their pointwise maximum $L_1(E)$. In this ideal construction, every true energy is a peak of height exactly $1$. Numbers are illustrative, not from a molecular calculation.}
\label{fig:mle_landscape}
\end{figure}

\paragraph{Phase-consistency landscape.} We represent the candidate energies of all phases by a continuous phase-consistency landscape over the appropriate energy window $\mathcal W$,
\begin{equation}
L_1(E) = \max_i \cos\!\big(\theta_i^{(1)} + E\tau_1\big), \qquad E\in\mathcal W.
\label{eq:landscape}
\end{equation}
For each phase $\theta_i^{(1)}$, $L_1=1$ at its candidate energies $E=(-\theta_i^{(1)}+2\pi k)/\tau_1$, where $\theta_i^{(1)}+E\tau_1$ is a multiple of $2\pi$. In the noiseless limit, with every physical root passing validation, these candidate energies include the true energies $E_i$ and their images $E_i+2\pi k/\tau_1$. With noisy phases, the peaks lie near rather than exactly at the true energies. Figure~\ref{fig:mle_landscape} illustrates the ideal construction for three schematic levels and a single $\tau_1$.

\begin{figure}
\centering
\includegraphics[width=\columnwidth]{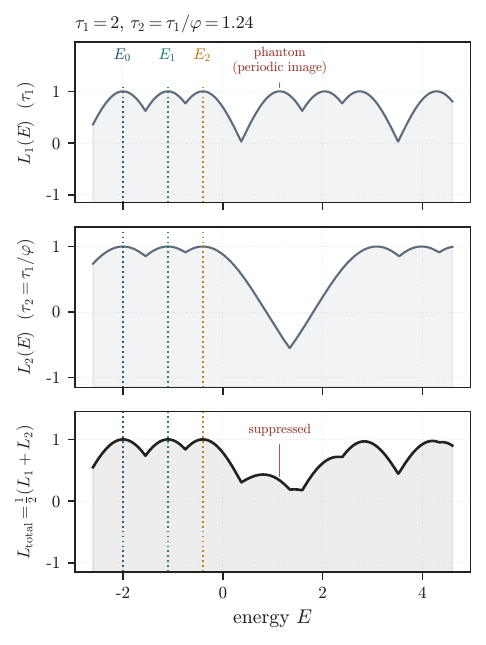}
\caption{Two runs at incommensurate time steps suppress a phantom peak in the ideal noiseless example. The energy window is widened past one period of $\tau_1$ so that a periodic image of $E_0$ appears in $L_1(E)$ (top). $L_2(E)$ uses $\tau_2=\tau_1/\varphi$ (middle), whose different periodicity does not reinforce that image. In the averaged landscape $L_{\mathrm{total}}$ (bottom), the true energies remain at height $1$ while the phantom is suppressed to $0.336$.}
\label{fig:incommensurate}
\end{figure}

\paragraph{Additional incommensurate time steps.} Besides the true energies $E_i$, $\mathcal W$ can contain their images $E_i+2\pi k/\tau_1$ with $k\neq0$: images of states whose true energies lie outside $\mathcal W$ and, if $\mathcal W$ is wider than $2\pi/\tau_1$, further images of states inside it. We call these images \emph{phantom peaks}. A phantom peak can be mistaken for a distinct state or can lie close to another true energy. If $\mathcal W$ is narrower than $2\pi/\tau_1$ and $L_1$ has exactly one peak above the acceptance level defined below for each target state, no further quantum resource is needed. Otherwise, we run one or more additional UQKSD calculations at incommensurate time steps $\tau_2,\tau_3,\dots$, obtain a landscape $L_g(E)$ from each via Eq.~\eqref{eq:landscape} with $\theta_i^{(g)}$ and $\tau_g$, and average them into the combined phase-consistency landscape,
\begin{equation}
L_{\mathrm{total}}(E) = \frac{1}{G}\sum_{g=1}^{G} L_g(E).
\label{eq:landscape_total}
\end{equation}
In the noiseless limit, with every physical root passing validation, a true energy is a peak of height $1$ in every $L_g$ and therefore in $L_{\mathrm{total}}$. A phantom peak of $L_1$ is generically not also a peak of an incommensurate $L_2$, so averaging lowers it. We take $\tau_2=\tau_1/\varphi$, with $\varphi=(1+\sqrt5)/2$ the golden ratio, which is poorly approximated by ratios of small integers. Images of the same state in the two landscapes therefore nearly coincide only at large winding integers: at the acceptance level $L_{\mathrm{total}}\geq0.9999$ used below, the first such coincidence lies 144 periods $2\pi/\tau_1$ away, farther than the energy range of all states with nonzero weight in $\ket{\phi_0}$ in Figs.~\ref{fig:noiseless_validation} and~\ref{fig:noisy_necessity}. An image $k$ periods away has $L_1=1$ but a phase misfit $2\pi\lVert k/\varphi\rVert$ in $L_2$, where $\lVert x\rVert$ is the distance from $x$ to the nearest integer; this misfit is $0.032$~rad at $k=89$, too large to pass the acceptance level, and first becomes small enough at $k=144$, where it is $0.0195$~rad. This argument does not exclude a peak formed by an image of one state in $L_1$ and an image of a different state in $L_2$. Because each landscape is a maximum over the phases of one run, a root of the base run never has to be paired with a root of the second run. The time step $\tau_2$ is fixed in advance, and only the decision to make the second run depends on the base run. Combining phases from two time steps is related to the Chinese remainder theorem, which is used for phase unwrapping in radar~\cite{Xia2007}. Here the ratio of the two time steps is irrational, and only energies inside $\mathcal W$ are considered. Figure~\ref{fig:incommensurate} demonstrates the suppression for the same ideal schematic system as Fig.~\ref{fig:mle_landscape}: at the true ground state, $L_1=L_2=L_{\mathrm{total}}=1$, while at the first periodic image of the ground state in $L_1$, $L_2=-0.328$ and $L_{\mathrm{total}}=0.336$.

\paragraph{Peak extraction and unwrapping.} We take the local maxima of $L_{\mathrm{total}}$ inside $\mathcal W$ with $L_{\mathrm{total}}\geq0.9999$ as guide energies $E_{\mathrm{guide},i}$. Noise shifts the phases of the two runs independently, so at a true energy the peaks of $L_1$ and $L_2$ no longer coincide exactly, and $L_{\mathrm{total}}$ falls below $1$ by an amount that grows with the difference between the two runs' energy errors. A phantom peak of one landscape, in contrast, is generically not a peak of the other and has a much lower $L_{\mathrm{total}}$ (Fig.~\ref{fig:incommensurate}). Since $L_{\mathrm{total}}=(L_1+L_2)/2$ and each $L_g$ is at most $1$, $L_{\mathrm{total}}\geq0.9999$ requires $L_g\geq0.9998$ in each landscape. By Eq.~\eqref{eq:landscape}, some phase of each run then satisfies $\cos(\theta_i^{(g)}+E\tau_g)\geq0.9998$, so $\theta_i^{(g)}+E\tau_g$ lies within $\arccos(0.9998)\approx0.02$~rad of a multiple of $2\pi$. The phases of both runs must therefore fit the peak energy to within $0.02$~rad. A lower level would accept more phantom peaks, and a higher one would reject true energies whose phases are noisier. Because our targets are the lowest states in $\mathcal W$, the guide energies are processed in ascending order, and if more guide energies than target states are accepted, only the lowest are kept. If fewer are accepted, each missing target is given the next unused base-run root, unwrapped to its image inside or nearest to $\mathcal W$. Each $E_{\mathrm{guide},i}$ is assigned the base-run phase that maximizes $\cos(\theta^{(1)}+E_{\mathrm{guide},i}\tau_1)$ among the phases not yet assigned, and we denote this phase by $\theta_i^{(1)}$, so each base-run phase is assigned to at most one guide energy. For each assigned pair, the integer winding number is
\begin{equation}
k_i = \mathrm{round}\!\left(\frac{E_{\mathrm{guide},i}-E_{\mathrm{aliased},i}^{(1)}}{2\pi/\tau_1}\right),
\label{eq:winding}
\end{equation}
where $E_{\mathrm{aliased},i}^{(1)}=-\theta_i^{(1)}/\tau_1$ is the raw, wrapped energy from the base run. Let $P=2\pi/\tau_1$ and $k_i^{\mathrm{true}}$ be the correct winding for the assigned root. Nearest-integer rounding is guaranteed to return it when
\begin{equation}
\left|E_{\mathrm{guide},i}-E_{\mathrm{aliased},i}^{(1)}-k_i^{\mathrm{true}}P\right|<\frac{P}{2}=\frac{\pi}{\tau_1}.
\label{eq:winding_condition}
\end{equation}
This bound includes both guide-energy error and base-phase error, and presumes that the intended root was assigned. A root misassignment lies outside the guarantee. The final, unwrapped energy is
\begin{equation}
E_{\mathrm{unwrapped},i} = E_{\mathrm{aliased},i}^{(1)} + k_i\,\frac{2\pi}{\tau_1}.
\label{eq:unwrap}
\end{equation}
The guide energy only needs to satisfy Eq.~\eqref{eq:winding_condition}, that is, lie within $\pi/\tau_1$ of $E_{\mathrm{aliased},i}^{(1)}+2\pi k_i^{\mathrm{true}}/\tau_1$. If the root assignment and $k_i$ are correct, Eq.~\eqref{eq:unwrap} adds an exact multiple of $2\pi/\tau_1$, so the unwrapped energy has the same error as the aliased energy $E_{\mathrm{aliased},i}^{(1)}=-\theta_i^{(1)}/\tau_1$: $E_{\mathrm{unwrapped},i}-E_i=E_{\mathrm{aliased},i}^{(1)}-(E_i-2\pi k_i/\tau_1)$, where $E_i-2\pi k_i/\tau_1$ is the exact energy moved into the same branch as $E_{\mathrm{aliased},i}^{(1)}$.

\section{Results and Discussion}
\label{sec:results}

\subsection{Simulation methodology and system set}
\label{sec:methodology}

\begin{table*}
\caption{Systems used in Sec.~\ref{sec:results}, all in the STO-3G basis, and the settings of Figs.~\ref{fig:noiseless_validation} and~\ref{fig:noisy_necessity}. $d$ is the bond length, the same for both bonds of H$_2$O and BeH$_2$ and for every neighboring pair in the H$_6$ chain. H$_2$O keeps a $104.5^\circ$ bond angle and BeH$_2$ is linear. $n_q$ is the number of qubits, $n_{\mathrm{ref}}$ the number of eigenstates with $|\langle E_n|\phi_0\rangle|^2>10^{-10}$, $|\gamma_0|^2$ the squared overlap of the Hartree--Fock reference with the exact ground state, $E_{\mathrm{gs}}$ the exact ground-state energy, and $\tau_{\mathrm{alias}}=\pi/|E_{\mathrm{gs}}|$ for the unshifted Hamiltonian [Eq.~\eqref{eq:tau_alias}] ($\tau_{\mathrm{alias}}$ and $\tau_{\max}$ are in atomic units). The windows $\mathcal W_{\mathrm{gs}}$ and $\mathcal W_{\mathrm{multi}}$ [Eqs.~\eqref{eq:window_gs} and~\eqref{eq:window_multi}] are given relative to $E_{\mathrm{HF}}$, and $\mathcal W_{\mathrm{multi}}$ is used in Fig.~\ref{fig:noiseless_validation} and for the first excited states in Fig.~\ref{fig:noisy_necessity}. For each figure, $M$ is the Krylov dimension. Figure~\ref{fig:noiseless_validation} targets the given number of lowest states at base time steps up to $\tau_{\max}$, and Fig.~\ref{fig:noisy_necessity} shows the given state.}
\begin{ruledtabular}
\begin{tabular}{lccccccccccccc}
 & & & & & & & & & \multicolumn{3}{c}{Fig.~\ref{fig:noiseless_validation}} & \multicolumn{2}{c}{Fig.~\ref{fig:noisy_necessity}} \\
\cline{10-12}\cline{13-14}
System & $d$ (\AA) & $n_q$ & $n_{\mathrm{ref}}$ & $|\gamma_0|^2$ & $E_{\mathrm{gs}}$ (Ha) & $\tau_{\mathrm{alias}}$ & $\mathcal W_{\mathrm{gs}}$ (Ha) & $\mathcal W_{\mathrm{multi}}$ (Ha) & $M$ & Targets & $\tau_{\max}$ & $M$ & State \\
\hline
H$_6$ chain & $1.5$ & 12 & 92 & $0.637$ & $-2.996$ & $1.049$ & $[-0.350,0]$ & $[-0.350,0.457]$ & 44 & 3 & $1.78$ & 44 & $E_1$ \\
HF & $1.5$ & 12 & 11 & $0.869$ & $-98.519$ & $0.032$ & $[-0.205,0]$ & $[-0.205,0.739]$ & 65 & 4 & $0.32$ & 65 & $E_1$ \\
N$_2$ & $1.5$ & 20 & 294 & $0.737$ & $-107.582$ & $0.029$ & $[-0.395,0]$ & $[-0.395,0.615]$ & 80 & 6 & $1.00$ & 20 & $E_1$ \\
LiH & $3.595$ & 12 & 29 & $0.597$ & $-7.787$ & $0.403$ & $[-0.234,0]$ & $[-0.234,0.290]$ & 38 & 4 & $0.89$ & 20 & $E_0$ \\
H$_2$O & $1.4$ & 14 & 69 & $0.858$ & $-74.910$ & $0.042$ & $[-0.238,0]$ & $[-0.238,0.776]$ & 67 & 5 & $1.05$ & 50 & $E_0$ \\
BeH$_2$ & $2.5$ & 14 & 64 & $0.543$ & $-15.352$ & $0.205$ & $[-0.281,0]$ & $[-0.281,0.418]$ & 80 & 6 & $1.06$ & 44 & $E_0$ \\
\end{tabular}
\end{ruledtabular}
\label{tab:systems}
\end{table*}

We validate the analysis of Sec.~\ref{sec:theory} with classical statevector simulations. Molecular Hamiltonians are built with OpenFermion~\cite{McClean2020} and PySCF~\cite{Sun2018,Sun2020} in the STO-3G minimal basis~\cite{Hehre1969}, mapped to qubits with the Jordan--Wigner transformation~\cite{Jordan1928}, and represented as sparse matrices. The reference state is the restricted Hartree--Fock determinant. $\mathbf S$ and $\mathbf F$ are assembled via Eqs.~\eqref{eq:S_toeplitz}--\eqref{eq:F_from_S} from the exact length-$(M+1)$ autocorrelation trajectory rather than from a Trotterized circuit, so that the behavior of the classical GEVP solver and the CURE-QKSD unwrapping is isolated from Trotter error, which has been characterized separately for UQKSD by Klymko and co-workers~\cite{Klymko2022}. In Trotterized Hubbard-model QKSD simulations, longer evolution resolves small gaps with fewer Krylov vectors but increases the accumulated Trotter error~\cite{Mirzakhani2026}. Exact propagation removes this error from our tests. For a time step $\tau$, the $M$ Krylov vectors span $T_{\mathrm{basis}}=(M-1)\tau$, and the additional autocorrelation value $c(M)$ needed for $\mathbf F$ is measured at $T_{\max}=M\tau$. We keep $M$ small, $M\leq80$, in Figs.~\ref{fig:noiseless_validation} and~\ref{fig:noisy_necessity}.

Sampling noise is modeled, as in Sec.~\ref{sec:illcond}, by perturbing each measured autocorrelation lag $c(k)$ with an independent complex Gaussian error of standard deviation $\sigma=10^{-4}$, then rebuilding $\widetilde{\mathbf S}$ and $\widetilde{\mathbf F}$ from the perturbed sequence via the same Toeplitz map used in the noiseless case [Eqs.~\eqref{eq:S_toeplitz}--\eqref{eq:F_from_S}]. This model contains only sampling noise. Under global depolarizing noise, for example, each measured $c(k)$ would also be damped by a factor that decays exponentially with the evolution time $k\tau$~\cite{OBrien2019,Ding2026}. In the noisy calculations we set the value of $\epsilon_{\mathrm{thresh}}$ to be of order $\sigma\sqrt M$. This is an estimate of $\lVert\Delta\mathbf S\rVert_2$: for the values of $M$ used here ($8$--$100$), our value of $\epsilon_{\mathrm{thresh}}$ is $0.6$--$0.9$ times the median $\lVert\Delta\mathbf S\rVert_2$. In the noiseless calculations, $\epsilon_{\mathrm{thresh}}=10^{-15}$, close to the double-precision roundoff of $\mathbf S$. Each noisy data point is the median error over $8$ noise realizations, and every realization returns an energy for every plotted state. Within a realization, the base run and any additional run receive independent noise, and plain UQKSD is applied to the same noisy base-run sequence as CURE-QKSD.

For CURE-QKSD, every system uses the root-validation settings $\delta=0.1$ and $c_{\mathrm{val}}=10^{-3}$ (Sec.~\ref{sec:cure_methodology}) and the window buffer $\delta_W=0.10$~Ha [Eqs.~\eqref{eq:window_gs}--\eqref{eq:window_multi}]. The uncorrected method in Fig.~\ref{fig:noisy_necessity}, which we call plain UQKSD, uses the same validation, reads every energy from the principal branch [Eq.~\eqref{eq:phase_extraction}], and reports the lowest energy as $E_0$ and the second lowest as $E_1$. Every CURE-QKSD result in this section uses at most one additional run, at $\tau_2=\tau_1/\varphi$, beyond the base run (two runs in total). We find the peaks of $L_{\mathrm{total}}$ by evaluating it at $4000$ equally spaced energies in $\mathcal W$, which lowers a peak's height by less than $10^{-7}$, far below the gap of $10^{-4}$ between $1$ and the acceptance level. When the second run is made, it adds the $M$ autocorrelation values $c(1),\dots,c(M)$ at the time step $\tau_2$. With the same number of measurements per value, this doubles the number of circuits and multiplies the total evolution time by $1+1/\varphi\approx1.6$.

Table~\ref{tab:systems} lists the six systems tested, their energy windows, and the settings of Figs.~\ref{fig:noiseless_validation} and~\ref{fig:noisy_necessity}. The systems span 12 to 20 qubits, Hartree--Fock ground-state overlaps from $54\%$ to $87\%$, and values of $\tau_{\mathrm{alias}}$ [Eq.~\eqref{eq:tau_alias}] over more than an order of magnitude. We use $\tau_{\mathrm{alias}}=\pi/|E_{\mathrm{gs}}|$ only to express the $\tau$ values of different systems on a common scale. An excited state $i$ is aliased for $\tau>\pi/|E_i|$, which differs from $\tau_{\mathrm{alias}}$. In these systems, $E_{\mathrm{CCSD}}$ lies within $14$~mHa of the exact ground-state energy, well inside the buffer, and every targeted excited state lies below the upper edge of $\mathcal W_{\mathrm{multi}}$.

Throughout this section, we label the eigenstates of $\hat H$ that have nonzero overlap with the Hartree--Fock reference, $|\langle E_n|\phi_0\rangle|^2>10^{-10}$, by increasing energy, $E_0\le E_1\le\cdots$, so that $E_0=E_{\mathrm{gs}}$, and call the state with energy $E_n$ the $n$th excited state. Only these eigenstates can be obtained, for any $M$ and $\tau$, because the Krylov vectors contain only the eigenstates present in $\ket{\phi_0}$. The reference $\ket{\phi_0}$ lies in a single symmetry sector of the electron-number, total-spin, and spatial-symmetry operators, and because $\hat H$ commutes with these operators, every Krylov vector stays in that sector. Eigenstates of $\hat H$ outside the sector therefore have zero overlap with the reference and with every Krylov vector, so the $n$th excited state defined here need not be the $n$th excited state of the full molecular Hamiltonian.

To find the cause of an inaccurate estimate, we compute the weight of each root on each exact eigenstate (Appendix~\ref{app:root_weights}). We call a state resolved when some root has at least $90\%$ of its weight on it.

\subsection{CURE-QKSD is essentially exact in the noiseless tests for large \texorpdfstring{$\tau$}{tau}}
\label{sec:results_noiseless}

\begin{figure*}
\centering
\includegraphics[width=\textwidth]{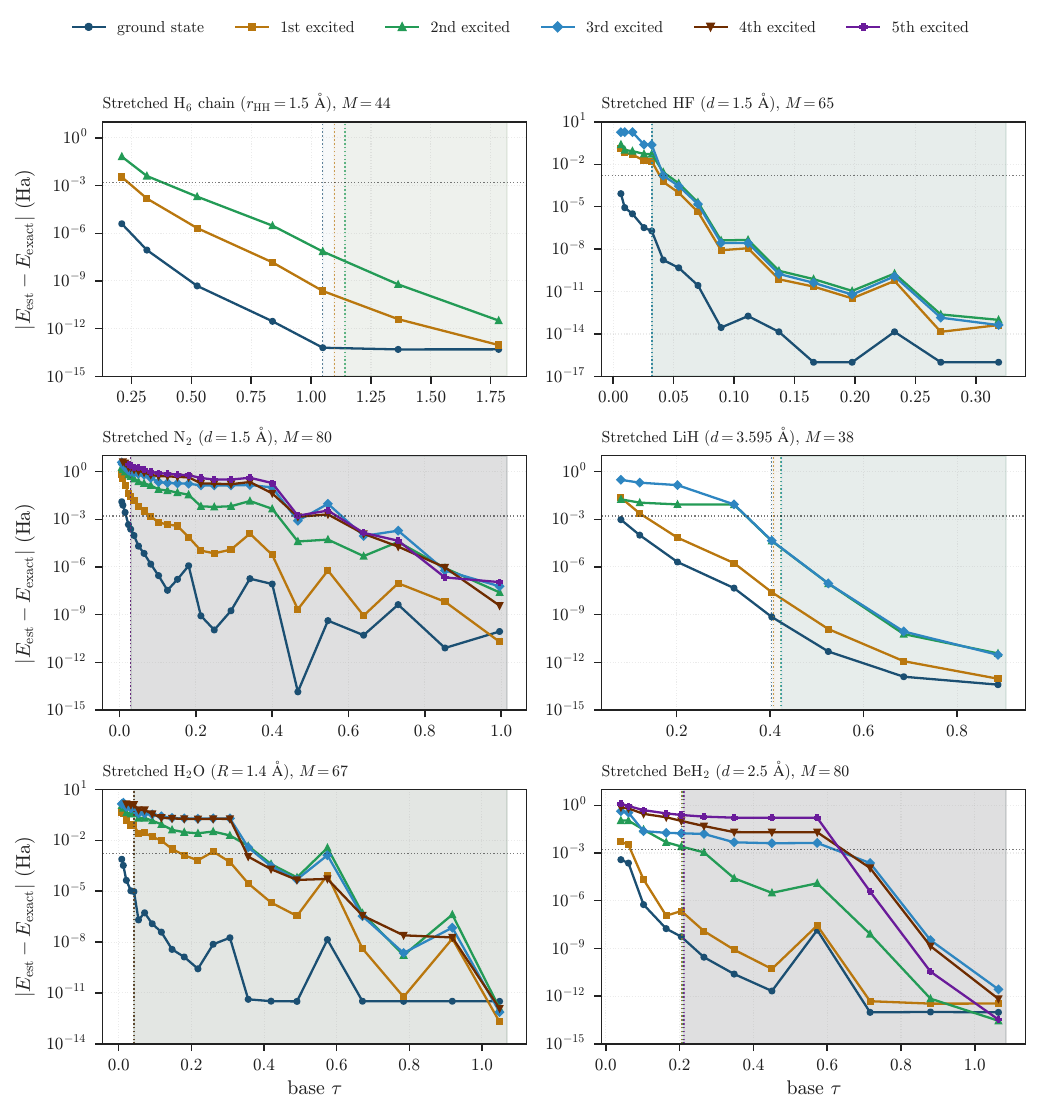}
\caption{Absolute error $|E_{\mathrm{est}}- E_{\mathrm{exact}}|$ of noiseless CURE-QKSD versus the base time step $\tau$ for the six systems of Table~\ref{tab:systems} (Krylov dimension $M$ given in each panel, numbers of targeted states in Table~\ref{tab:systems}). The lowest returned energy is compared with $E_0$, the second lowest with $E_1$, and so on. Appendix~\ref{app:root_weights} gives the cause of each error above chemical accuracy (horizontal dotted line). Shading marks each state's aliased region, $\tau>\pi/|E_i|$.}
\label{fig:noiseless_validation}
\end{figure*}

Figure~\ref{fig:noiseless_validation} shows the noiseless error of every targeted state versus the base time step $\tau$. All systems use the same set of base time steps, starting at $0.2\,\tau_{\mathrm{alias}}$, and every time step up to $\tau_{\max}$ of Table~\ref{tab:systems} is plotted. The lowest returned energy is the estimate of $E_0$, the second lowest that of $E_1$, and so on, so a state that is not resolved also affects the estimates of the states above it. At $\tau_{\max}$, every targeted energy has an error below $10^{-10}$~Ha in the 12- and 14-qubit systems and between $2\times10^{-11}$ and $1.1\times10^{-7}$~Ha for N$_2$. The error rises sharply at a few isolated time steps, mainly because of aliasing in the Krylov basis (Appendix~\ref{app:basis_alias}).

Figure~\ref{fig:noiseless_validation} has 443 points, one for each system, time step, and targeted state, and 237 of them are within chemical accuracy. Every error above chemical accuracy comes from the base run: either the state or a lower one is not resolved, so the estimate is a root shared by several states or the root of another state, or the state is resolved but its root is still inaccurate. 
None of the errors is caused by a step of CURE-QKSD: no resolved state loses its root in validation, lies outside $\mathcal W$, or falls below the acceptance level, and every winding number is correct (Appendix~\ref{app:root_weights}). The second run does not change any result in Fig.~\ref{fig:noiseless_validation}.

Nine targeted excited states have a reference weight below $1\%$: $0.56\%$ for $E_2$ of H$_6$, $0.98\%$ for $E_2$ of BeH$_2$, $0.51\%$, $0.22\%$, and $0.47\%$ for $E_1$--$E_3$ of H$_2$O, and $0.57\%$, $0.14\%$, $0.84\%$, and $0.92\%$ for $E_2$--$E_5$ of N$_2$. Their errors at $\tau_{\max}$ are also below $10^{-10}$~Ha, or $1.1\times10^{-7}$~Ha for N$_2$. The H$_2$O states show how the time step needed to resolve a state depends on its distance to its neighbors. $E_2$ and $E_3$ are only $38$~mHa apart, whereas the other consecutive states are $81$--$332$~mHa apart, and resolving them requires a total evolution time $T_{\mathrm{basis}}$ of order $1/(38~\mathrm{mHa})=26$~a.u.\ (Sec.~\ref{sec:illcond}). At $M=67$, they are not resolved below $8.5\,\tau_{\mathrm{alias}}$, where the estimates of $E_2$--$E_4$ have errors of $0.02$--$1.6$~Ha. At $8.5\,\tau_{\mathrm{alias}}$, where $T_{\mathrm{basis}}=(M-1)\tau=24$~a.u., both are resolved, and from there on all five errors are below $5$~mHa. With an unaliased time step, $\tau\leq\tau_{\mathrm{alias}}$, the same $T_{\mathrm{basis}}$ would require $M\geq562$. 

\subsection{CURE-QKSD reaches chemical accuracy in the noisy aliased regime}
\label{sec:results_noisy}

\begin{figure*}
\centering
\includegraphics[width=\textwidth]{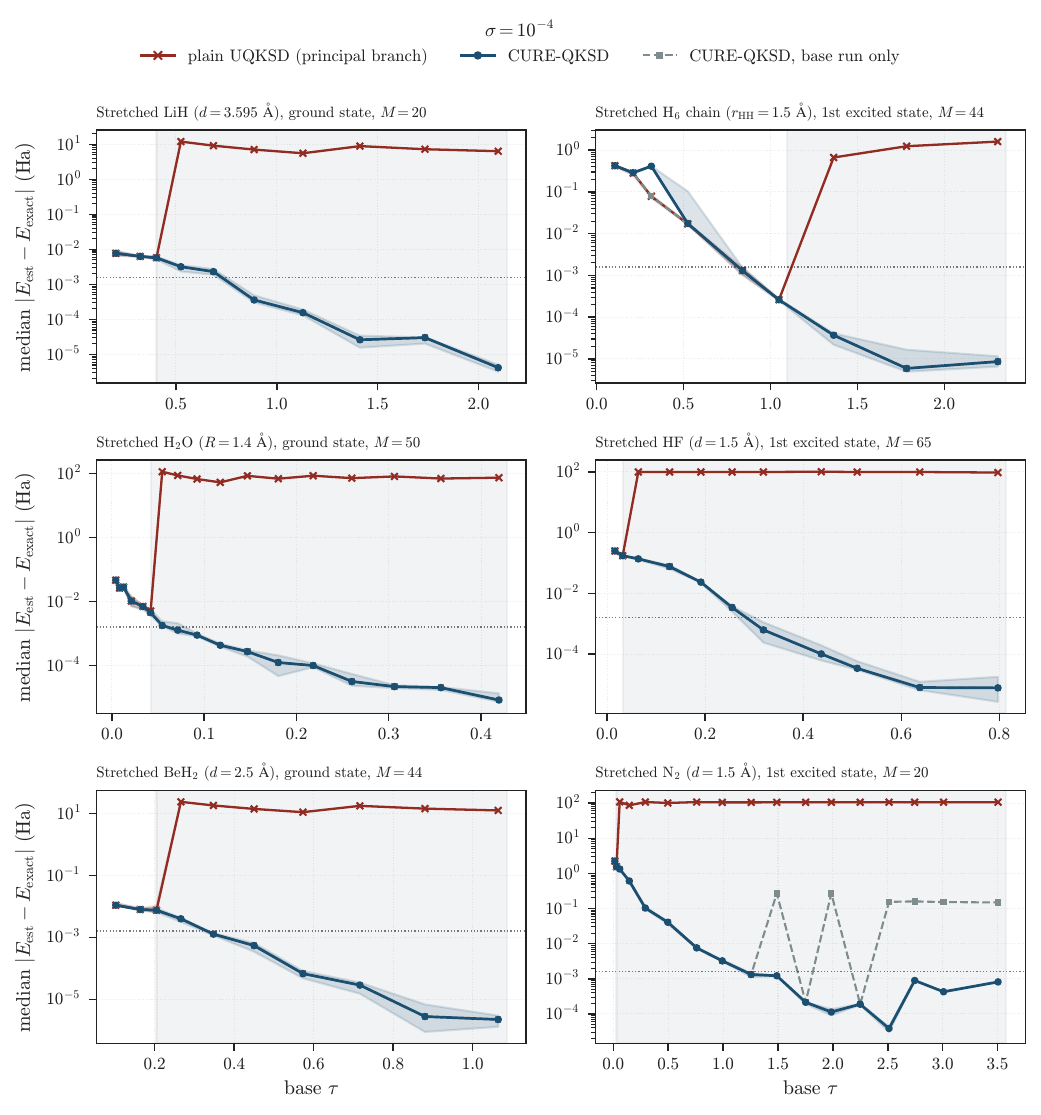}
\caption{Median error (8 noise realizations, $\sigma=10^{-4}$) versus $\tau$ for the ground state of LiH, H$_2$O, and BeH$_2$ (left) and the first excited state of H$_6$, HF, and N$_2$ (right): plain UQKSD (red), CURE-QKSD (blue), and CURE-QKSD with the base run only (gray, dashed). Shading marks the aliased region of the plotted state, $\tau>\pi/|E_i|$. The base-run-only curve differs from CURE-QKSD only for the unresolved first excited state of H$_6$ at $0.2$--$0.5\,\tau_{\mathrm{alias}}$ and for N$_2$ at $51$, $68$, and $86$--$120\,\tau_{\mathrm{alias}}$, where the base run alone accepts a phantom peak (Sec.~\ref{sec:results_noisy}).}
\label{fig:noisy_necessity}
\end{figure*}

Figure~\ref{fig:noisy_necessity} repeats the $\tau$ sweep with noise, $\sigma=10^{-4}$, and compares CURE-QKSD with plain UQKSD for the ground states of LiH, H$_2$O, and BeH$_2$ and for the first excited states of H$_6$, HF, and N$_2$, whose reference weights are $2.5\%$, $2.2\%$, and $8.4\%$. For $E_1$, both methods report their second-lowest energy, and CURE-QKSD uses $\mathcal W_{\mathrm{multi}}$. H$_6$ and HF use the $M$ of Fig.~\ref{fig:noiseless_validation}, and the other four systems use a smaller $M$ (Table~\ref{tab:systems}).

With noise, the base run does not resolve $E_1$ in any realization up to $0.5\,\tau_{\mathrm{alias}}$ for H$_6$, $6\,\tau_{\mathrm{alias}}$ for HF, and $17\,\tau_{\mathrm{alias}}$ for N$_2$ (Appendix~\ref{app:root_weights}). At these time steps, the energy reported as $E_1$ comes from a root that mixes $E_1$ with other states, and its median error is $0.017$--$2$~Ha. Without noise, a shorter $T_{\mathrm{basis}}=(M-1)\tau$ suffices at the same $M$: $E_1$ of H$_6$ is resolved from $0.2\,\tau_{\mathrm{alias}}$ and that of HF from $\tau_{\mathrm{alias}}$, so noise increases the $T_{\mathrm{basis}}$ needed to resolve $E_1$. For N$_2$ at $M=20$, $E_1$ is not resolved below $17\,\tau_{\mathrm{alias}}$ even without noise. Once $E_1$ is resolved, the reported root has its largest weight on $E_1$ in every realization. For the ground states, apart from two realizations for BeH$_2$ at $\tau_{\mathrm{alias}}$, the reported root has its largest weight on $E_0$, so the error comes from the weight of this root on other states.

Plain UQKSD fails as soon as the plotted state is aliased, whereas the CURE-QKSD error keeps decreasing. Each panel of Fig.~\ref{fig:noisy_necessity} plots one state, $E_0$ or $E_1$, which is aliased for $\tau>\pi/|E_i|$ (shaded). Below this time step, the two methods return the same energy in every realization, except for the unresolved $E_1$ of H$_6$ at $0.2$--$0.5\,\tau_{\mathrm{alias}}$, where the second run makes CURE-QKSD report a different mixed root; neither energy is accurate, and the difference is not caused by aliasing. For $\tau>\pi/|E_i|$, the median error of plain UQKSD is $0.7$--$115$~Ha at every $\tau$ tested, because each energy is returned only modulo $2\pi/\tau$. The CURE-QKSD error shows no jump at $\pi/|E_i|$, indicating that CURE-QKSD has correctly identified the winding number. The ground states reach chemical accuracy by $1.7$--$2.2\,\tau_{\mathrm{alias}}$ and the first excited states by $0.8$ (H$_6$), $10$ (HF), and $43\,\tau_{\mathrm{alias}}$ (N$_2$). At fixed $\tau$, the ground-state error also decreases faster with $M$ at aliased time steps than at $0.5\,\tau_{\mathrm{alias}}$, and reaches chemical accuracy at smaller $M$ (Appendix~\ref{app:M_sweep}). At larger $\tau$, the median errors decrease further, with fluctuations, to minima of $2\times10^{-6}$--$4\times10^{-5}$~Ha. For N$_2$, the minimum is at $86\,\tau_{\mathrm{alias}}$, and the error rises again to $4$--$9\times10^{-4}$~Ha at $94$--$120\,\tau_{\mathrm{alias}}$.

Overall, we find that most of the systems converge with only a single CURE-QKSD run. 
The second run changes a result only for the unresolved $E_1$ of H$_6$ at $0.2$--$0.5\,\tau_{\mathrm{alias}}$ (previous paragraph) and for $E_1$ of N$_2$ at $51$, $68$, and $86$--$120\,\tau_{\mathrm{alias}}$ ($\tau=1.49$--$3.50$). At these N$_2$ time steps, the base run alone returns the energy of another root as $E_1$ in six to eight of the eight realizations, with median errors of $0.15$--$0.27$~Ha: a phantom peak, the image of a higher state, lies in $\mathcal W_{\mathrm{multi}}$ below $E_1$ (below $E_0$ at $51$ and $68\,\tau_{\mathrm{alias}}$) and is accepted, because with one run every image of a root has $L_1=1$. With the second run, $L_{\mathrm{total}}$ of this peak is $0.983$--$0.999$, below the acceptance level, and CURE-QKSD returns $E_1$ within chemical accuracy in every realization. These time steps matter for N$_2$: at $M=20$, $E_1$ reaches chemical accuracy only at $43\,\tau_{\mathrm{alias}}$, and its median error is smallest, $3.9\times10^{-5}$~Ha, at $86\,\tau_{\mathrm{alias}}$, where the base run alone fails.

\subsection{Shifting the Hamiltonian does not replace CURE-QKSD}
\label{sec:results_shift}

The results above show that CURE-QKSD corrects aliasing at large time steps. Each of those calculations used the bare Hamiltonian. A simpler way to delay aliasing is to shift the Hamiltonian by a classical estimate $E_s$ of the target energy, $\hat H\to\hat H-E_s\mathbf 1$, so that state $i$ is aliased only for $\tau>\pi/|E_i-E_s|$. The shift needs no new measurements: multiplying each $c(k)$ by $e^{ikE_s\tau}$ leaves the roots unchanged apart from a factor $e^{iE_s\tau}$, so each energy is read from the branch $[E_s-\pi/\tau,E_s+\pi/\tau)$ instead of $[-\pi/\tau,\pi/\tau)$. For $E_s=E_{\mathrm{HF}}$, the ground-state phase depends only on the usually small correlation energy $E_{\mathrm{gs}}-E_{\mathrm{HF}}$, and the ground state stays in the branch up to a much larger $\tau$.

The shift, however, applies to every root, not only to the target. All energies are read from one branch of width $2\pi/\tau$ centered at $E_s$. An energy above this branch wraps around and appears $2\pi/\tau$ lower, where it can fall below $E_0$ or $E_1$ and is then reported in their place. The shift therefore works only if all these energies fit inside the branch. This becomes harder as $\tau$ grows and the branch narrows, and once the branch is narrower than the spread of the energies, no choice of $E_s$ works.

\begin{figure*}
\centering
\includegraphics[width=\textwidth]{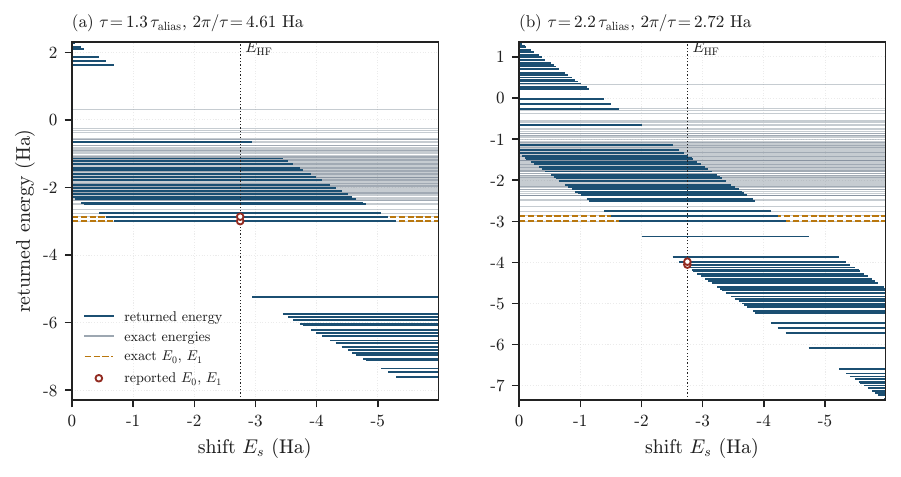}
\caption{Energies returned by plain UQKSD of $\hat H-E_s\mathbf 1$ versus the shift $E_s$ for H$_6$ at (a) $\tau=1.3\,\tau_{\mathrm{alias}}$ and (b) $2.2\,\tau_{\mathrm{alias}}$ ($M=44$, first noise realization of Fig.~\ref{fig:noisy_necessity}). Each root is read from the branch $[E_s-\pi/\tau,E_s+\pi/\tau)$, so each of its images is returned over a range of $E_s$ of width $2\pi/\tau$ (blue segments). Gray lines mark the exact energies of the $n_{\mathrm{ref}}=92$ eigenstates with $|\langle E_n|\phi_0\rangle|^2>10^{-10}$, dashed lines $E_0$ and $E_1$, and the dotted line $E_s=E_{\mathrm{HF}}$. Circles mark the two lowest energies at $E_s=E_{\mathrm{HF}}$, which the shifted calculation reports as $E_0$ and $E_1$.}
\label{fig:hf_shift}
\end{figure*}

Figure~\ref{fig:hf_shift} shows this for H$_6$, plotting the energies returned at each $E_s$ for $\tau=1.3$ and $2.2\,\tau_{\mathrm{alias}}$. At $1.3\,\tau_{\mathrm{alias}}$ [Fig.~\ref{fig:hf_shift}(a)], the branch width $2\pi/\tau=4.61$~Ha exceeds the spread of the energies, $2.35$~Ha, and the shift by $E_{\mathrm{HF}}$ returns $E_0$ and $E_1$ within chemical accuracy. At $2.2\,\tau_{\mathrm{alias}}$ [Fig.~\ref{fig:hf_shift}(b)], the branch width is $2.72$~Ha. At $E_s=E_{\mathrm{HF}}$, images of higher states wrap below $E_0$ and are reported as $E_0$ and $E_1$, with errors of $1.1$~Ha, even though $E_0$ and $E_1$ themselves are still in the branch. A shift of about $-1.9$~Ha, $0.85$~Ha above $E_{\mathrm{HF}}$, would return them correctly, but choosing it requires knowing where the higher energies lie, and these are themselves aliased.

The shift fails in the same way at other time steps and in other systems. For H$_6$, its errors are $0.13$--$1.2$~Ha at $1.7$ and $2.2\,\tau_{\mathrm{alias}}$, whereas those of CURE-QKSD are below $10^{-5}$~Ha. For HF and N$_2$, the shift matches CURE-QKSD as long as $\pi/\tau$ is larger than the distance from $E_{\mathrm{HF}}$ to the highest state that passes validation, $2.60$ and $3.85$~Ha, that is, up to $25$ and $26\,\tau_{\mathrm{alias}}$. At every larger $\tau$ tested, its errors are $0.56$--$2.7$~Ha, whereas those of CURE-QKSD are at most $4.3\times10^{-6}$~Ha for HF and $3\times10^{-3}$~Ha for N$_2$. In all three systems, $E_0$ and $E_1$ remain in the branch centered at $E_{\mathrm{HF}}$, and the shift fails only because images of higher roots fall below them. The time step up to which the shift works is therefore set by this highest state in the energy spectrum rather than by the target, so the shift cannot replace the per-root winding correction of CURE-QKSD.

\section{Conclusion}
\label{sec:conclusion}

In this paper, we addressed a tension in the choice of time step for unitary QKSD. A small time step keeps the target energies inside the principal branch but makes consecutive Krylov vectors nearly parallel and the overlap matrix ill-conditioned. A large time step improves the conditioning but aliases the energies, which are then known only modulo $2\pi/\tau$. We resolved this tension with CURE-QKSD, a classical post-processing step that recovers the missing multiples of $2\pi/\tau$, so that large, well-conditioned time steps can be used.

The small-step side of this tension is more serious than ill-conditioning alone. Resolving closely spaced energies requires a long total evolution time, $T_{\mathrm{basis}}=(M-1)\tau$, and we showed in a toy model that reaching it with many closely spaced Krylov vectors does not help (Sec.~\ref{sec:illcond}). At fixed $T_{\mathrm{basis}}$, the number of eigenvalues of $\mathbf S$ above the noise level $\lVert\Delta\mathbf S\rVert_2$ stops growing at a value of $M$ set by $T_{\mathrm{basis}}$, and the smallest error attainable by thresholding stops decreasing at the same $M$. Beyond that point, a cutoff slightly below the noise level can select a spurious root and raise the error by orders of magnitude. The extra vectors therefore add cost and a failure mode, but no accuracy.

CURE-QKSD leaves the UQKSD circuits and GEVP unchanged. A window built from classical estimates, here the CCSD and Hartree--Fock energies and the HOMO--LUMO gap, limits the candidate energies of each root to a finite set, and a phase-consistency landscape over that window marks each candidate as a peak. When the window also contains images of states, a second run at an incommensurate time step suppresses them. The winding number found this way is applied to the base-run phase, so if it is correct, the reported energy has exactly the base-run error. Phase-estimation algorithms resolve the same ambiguity with a sequence of evolution times that starts from an unaliased one, but CURE-QKSD takes that coarse information from the classical window instead, so every run uses a large time step. The method has three limitations. The target energies must lie inside the window, which can fail if the classical estimates are poor. The acceptance test does not exclude a peak formed by images of two different states. CURE-QKSD also has no proven error bound of the kind available for phase-estimation algorithms, and a comparison with them at equal quantum resources remains to be done.

Other approaches address the conditioning side of the tension by changing the basis. Orthogonal quantum Krylov diagonalization builds Lanczos polynomials of $\hat H$ with block encodings and generalized quantum signal processing, which removes the need to threshold~\cite{Rammal2026}, and quantum Krylov using unitary decomposition replaces time evolution by powers of a tunable Hermitian function of $\hat H$~\cite{Asthana2025}. Both change the quantum circuits. CURE-QKSD keeps the UQKSD basis and its autocorrelation-only measurements, and changes only the classical step that reads energies from the roots.

In noiseless tests on six molecular systems with 12--20 qubits, every targeted energy at the largest time step is recovered to within $10^{-10}$~Ha, or $1.1\times10^{-7}$~Ha for N$_2$, and every winding number is correct. In the same tests, two H$_2$O states $38$~mHa apart are resolved at $M=67$ with $\tau=8.5\,\tau_{\mathrm{alias}}$, whereas an unaliased time step would require $M\geq562$. With Gaussian noise of $\sigma=10^{-4}$ on the autocorrelations, the ground states of LiH, H$_2$O, and BeH$_2$ and the first excited states of H$_6$, HF, and N$_2$ are within chemical accuracy at aliased time steps with $M=20$--$65$. At these time steps, the principal-branch energies of the unshifted Hamiltonian are wrong by $0.7$--$115$~Ha. A shift by the Hartree--Fock energy delays this failure but does not remove it: once the highest state that passes validation lies more than $\pi/\tau$ above $E_{\mathrm{HF}}$, images of higher states fall below the targets and are reported in their place. Large, aliased time steps are therefore usable in unitary QKSD, provided that the classical window contains the targets and the phase ambiguity is resolved in post-processing.

\section{Acknowledgments} 
The authors are grateful for the generous support from the  Department of Energy: DE-SC0025430.

\bibliographystyle{apsrev4-2}
\bibliography{references}
\clearpage
\appendix
\section{Eigenvalue perturbation bound}
\label{app:weyl}
We order the eigenvalues of an $M\times M$ Hermitian matrix $\mathbf X$ as $\lambda_1(\mathbf X)\ge\cdots\ge\lambda_M(\mathbf X)$. For Hermitian $\mathbf A$ and $\mathbf B$, Weyl's inequalities relate the eigenvalues of $\mathbf A+\mathbf B$ directly to those of $\mathbf A$ and $\mathbf B$~\cite{HornJohnson2013}:
\begin{align}
\lambda_{j+k-1}(\mathbf A+\mathbf B)&\le\lambda_j(\mathbf A)+\lambda_k(\mathbf B),\label{eq:weyl_upper}\\
\lambda_j(\mathbf A)+\lambda_k(\mathbf B)&\le\lambda_{j+k-M}(\mathbf A+\mathbf B),\label{eq:weyl_lower}
\end{align}
where Eq.~\eqref{eq:weyl_upper} holds for $j+k-1\le M$ and Eq.~\eqref{eq:weyl_lower} for $j+k-M\ge1$.

We take $\mathbf A=\mathbf S$ and $\mathbf B=\Delta\mathbf S$, so that $\mathbf A+\mathbf B=\widetilde{\mathbf S}$. For $k=1$, the condition on Eq.~\eqref{eq:weyl_upper} reduces to $j\le M$, so for every $j$
\begin{equation}
\lambda_j(\widetilde{\mathbf S})\le\lambda_j(\mathbf S)+\lambda_1(\Delta\mathbf S).
\label{eq:app_upper}
\end{equation}
For $k=M$, the condition on Eq.~\eqref{eq:weyl_lower} reduces to $j\ge1$, so for every $j$
\begin{equation}
\lambda_j(\mathbf S)+\lambda_M(\Delta\mathbf S)\le\lambda_j(\widetilde{\mathbf S}).
\label{eq:app_lower}
\end{equation}
Subtracting $\lambda_j(\mathbf S)$ from Eqs.~\eqref{eq:app_upper} and \eqref{eq:app_lower} gives
\begin{equation}
\lambda_M(\Delta\mathbf S)\le\lambda_j(\widetilde{\mathbf S})-\lambda_j(\mathbf S)\le\lambda_1(\Delta\mathbf S).
\label{eq:weyl_two_sided}
\end{equation}
Because $\Delta\mathbf S$ is Hermitian, its singular values are the absolute values of its eigenvalues. Every eigenvalue lies between $\lambda_M(\Delta\mathbf S)$ and $\lambda_1(\Delta\mathbf S)$, so the largest absolute value occurs at one of these two ends, and
\begin{equation}
\lVert\Delta\mathbf S\rVert_2=\max\{|\lambda_1(\Delta\mathbf S)|,|\lambda_M(\Delta\mathbf S)|\}.
\end{equation}
Hence $\lambda_1(\Delta\mathbf S)\le\lVert\Delta\mathbf S\rVert_2$ and $\lambda_M(\Delta\mathbf S)\ge-\lVert\Delta\mathbf S\rVert_2$, and Eq.~\eqref{eq:weyl_two_sided} gives $-\lVert\Delta\mathbf S\rVert_2\le\lambda_j(\widetilde{\mathbf S})-\lambda_j(\mathbf S)\le\lVert\Delta\mathbf S\rVert_2$, which is Eq.~\eqref{eq:weyl_overlap}.

\section{Ground-state error versus Krylov dimension}
\label{app:M_sweep}

\begin{figure*}
\centering
\includegraphics[width=\textwidth]{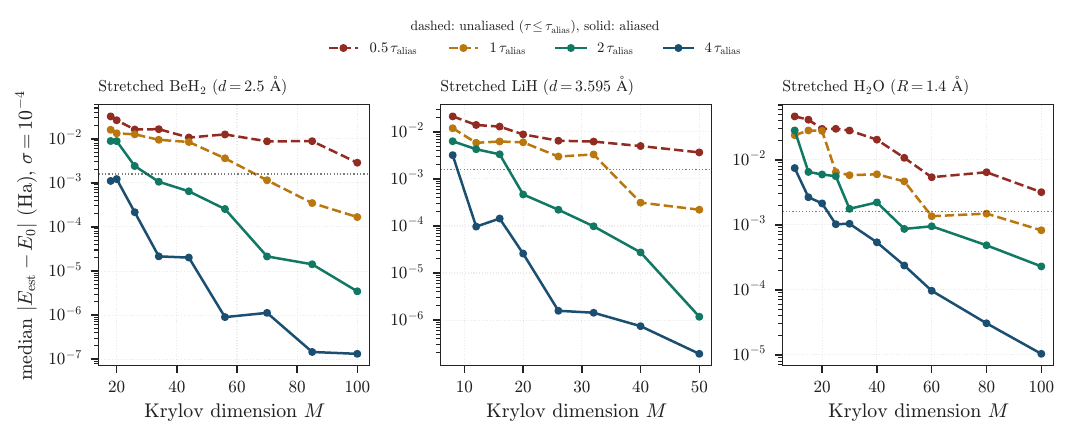}
\caption{Median ground-state error (8 noise realizations, $\sigma=10^{-4}$) versus Krylov dimension $M$ at four fixed ratios $\tau/\tau_{\mathrm{alias}}$, for BeH$_2$, LiH, and H$_2$O at the bond lengths of Table~\ref{tab:systems}. Dashed: $\tau\le\tau_{\mathrm{alias}}$; solid: aliased; dotted: chemical accuracy.}
\label{fig:M_sweep_noisy}
\end{figure*}

Figure~\ref{fig:M_sweep_noisy} shows how the ground-state error depends on $M$ at fixed $\tau$. For BeH$_2$, LiH, and H$_2$O we fix $\tau$ at $0.5$, $1$, $2$, and $4\,\tau_{\mathrm{alias}}$ (dashed: $\tau\le\tau_{\mathrm{alias}}$; solid: aliased) and compute the median ground-state error as a function of $M$ over 8 noise realizations with $\sigma=10^{-4}$. At $0.5\,\tau_{\mathrm{alias}}$, the error decreases by a factor of only $6$--$15$ over the range of $M$ tested and stays above chemical accuracy: from $3.2\times10^{-2}$ to $2.9\times10^{-3}$~Ha between $M=18$ and $100$ for BeH$_2$, from $2.1\times10^{-2}$ to $3.7\times10^{-3}$~Ha between $M=8$ and $50$ for LiH, and from $4.7\times10^{-2}$ to $3.2\times10^{-3}$~Ha between $M=10$ and $100$ for H$_2$O. At $2$ and $4\,\tau_{\mathrm{alias}}$, it decreases by factors of $120$--$17\,000$ over the same ranges. BeH$_2$ at $2\,\tau_{\mathrm{alias}}$ crosses chemical accuracy between $M=26$ and $34$ and reaches $3.5\times10^{-6}$~Ha at $M=100$, and at $4\,\tau_{\mathrm{alias}}$ it is already below chemical accuracy at the smallest $M$ tested ($1.1\times10^{-3}$~Ha at $M=18$). LiH crosses chemical accuracy between $M=16$ and $20$ at $2\,\tau_{\mathrm{alias}}$ and is below it from $M=12$ at $4\,\tau_{\mathrm{alias}}$. H$_2$O crosses it between $M=40$ and $50$ at $2\,\tau_{\mathrm{alias}}$, but between $M=20$ and $25$ at $4\,\tau_{\mathrm{alias}}$, and reaches $1.0\times10^{-5}$~Ha at $M=100$. The larger, aliased time steps therefore reach chemical accuracy at smaller $M$, and CURE-QKSD makes them usable by determining the branch of each energy.

\section{Aliasing in the Krylov basis}
\label{app:basis_alias}
Expanding the reference state as $\ket{\phi_0}=\sum_\ell\gamma_\ell\ket{E_\ell}$, with weights $w_\ell=|\gamma_\ell|^2$ and phases $z_\ell=e^{-iE_\ell\tau}$, gives $\ket{\phi_n}=\sum_\ell\gamma_\ell z_\ell^{\,n}\ket{E_\ell}$. The UQKSD matrices of Eqs.~\eqref{eq:S_toeplitz} and \eqref{eq:F_from_S} then factor as
\begin{equation}
\mathbf S=\mathbf V^\dagger\mathbf D\mathbf V,\qquad
\mathbf F=\mathbf V^\dagger\mathbf D\mathbf Z\mathbf V,
\label{eq:vandermonde}
\end{equation}
where $V_{\ell n}=z_\ell^{\,n}$ is a Vandermonde matrix with one row per eigenstate, $\mathbf D=\mathrm{diag}(w_\ell)$, and $\mathbf Z=\mathrm{diag}(z_\ell)$. The energies enter $\mathbf S$ and $\mathbf F$ only through $z_\ell$, that is, only modulo $2\pi/\tau$.

If $(E_i-E_j)\tau=2\pi m$ for a nonzero integer $m$, then $z_i=z_j$ and rows $i$ and $j$ of $\mathbf V$ are identical. Every Krylov vector then contains the two states only in the fixed combination $\gamma_i\ket{E_i}+\gamma_j\ket{E_j}$, and $\mathbf S$ and $\mathbf F$ are exactly those of a single state with phase $z_i$ and weight $w_i+w_j$. The Krylov vectors can still be linearly independent, because the other states contribute distinct phases, but no vector in $\mathcal K_M$ separates $\ket{E_i}$ from $\ket{E_j}$, and the GEVP can return at most one root for the two states. Since the data are identical to those of the merged state, no post-processing of $\mathbf S$ and $\mathbf F$ can recover the missing state. Shen \textit{et al.} pointed out that small steps make the Krylov vectors nearly linearly dependent, while large steps with $(E_j-E_0)\tau\ge2\pi$ wrap the phases around the unit circle and make such degeneracies possible~\cite{Shen2023}.

Near coincidences can be treated in the same way. Let $\delta_{ij}\in(-\pi,\pi]$ be $(E_i-E_j)\tau$ reduced modulo $2\pi$, that is, the separation of $z_i$ and $z_j$ on the unit circle. Shifting $E_j$ by a multiple of $2\pi/\tau$ does not change $z_j$, so by Eq.~\eqref{eq:vandermonde} $\mathbf S$ and $\mathbf F$ are unchanged if $E_j$ is replaced by its image $E_i-\delta_{ij}/\tau$. The basis therefore sees the two states as if they were separated by the folded gap $|\delta_{ij}|/\tau$ instead of $|E_i-E_j|$, which can be much larger. The resolution requirement of Sec.~\ref{sec:illcond} then applies to the folded gap, $(M-1)|\delta_{ij}|=T_{\mathrm{basis}}|\delta_{ij}|/\tau\gtrsim1$. Each time step also adds only the relative phase $\delta_{ij}$ between the two states, so, following the estimate of Klymko and co-workers, the pair is usefully resolved only if $|\delta_{ij}|\gtrsim\eta_{\mathrm{noise}}$~\cite{Klymko2022}. The effect of noise is clearest in a system containing only these two states. There $\mathbf S$ has two nonzero eigenvalues, and the smaller one, which belongs to the direction needed to separate the two states, is approximately $\bar w\,M(M^2-1)\delta_{ij}^2/12$, with $\bar w=w_iw_j/(w_i+w_j)$. It decreases as $\delta_{ij}^2$, so at fixed noise it falls below $\lVert\Delta\mathbf S\rVert_2$ as $\delta_{ij}\to0$. A threshold set at the noise level then removes it, and the basis can no longer separate the two states.

It remains to ask which time steps avoid coincidences. Let all states with non-negligible weight lie in $[E_0,E_{\mathrm{top}}]$, and let $R=E_{\mathrm{top}}-E_0$. An exact coincidence with $m\neq0$ requires two phases to differ by at least a full turn, $|E_i-E_j|\tau\ge2\pi$, so it cannot occur if $R\tau<2\pi$. This condition is sufficient but not necessary. For a discrete spectrum, two phases coincide exactly only at special values of $\tau$, so a generic $\tau$ gives distinct phases even when $R\tau>2\pi$. Keeping every gap at its true value needs a stronger condition. Because $\delta_{ij}$ is measured the shorter way around the circle, a pair with $\pi<|E_i-E_j|\tau<2\pi$ is seen at the smaller separation $2\pi-|E_i-E_j|\tau$ even when $R\tau<2\pi$. Every folded gap therefore equals the true gap only if all phases fit in half a circle, $R\tau\le\pi$. Error analyses of real-time QKSD use $\tau=\pi/R$, with $E_{\mathrm{top}}$ either the largest eigenvalue~\cite{Kirby2024} or a chosen cutoff, the states above it entering the error bound through their weights~\cite{Epperly2022}.

These coincidences are a property of a basis built from one uniform step, not of the post-processing. UQKSD and HQKSD at the same $\tau$ use the same basis and therefore have the same coincidences, and unwrapping the GEVP eigenvalues neither creates nor removes them.

A possible solution for this problem is to add Krylov vectors at a second step $\tau'$ such that $\tau'/\tau$ is irrational. If a nonzero gap satisfies $(E_i-E_j)\tau\in2\pi\mathbb Z$, it cannot also satisfy $(E_i-E_j)\tau'\in2\pi\mathbb Z$, so the enlarged basis contains both states. Unlike CURE-QKSD, this technique changes the basis itself. Combining vectors generated with $\tau$ and $\tau'$ requires additional overlaps at mixed times $n\tau'-m\tau$, and therefore additional quantum measurements rather than only classical post-processing.

\section{Root weights and the causes of the errors}
\label{app:root_weights}

For each root, we form its Ritz vector, $\ket{\Psi}=\sum_{n=0}^{M-1}c_n\ket{\phi_n}$, where $\mathbf c$ is the eigenvector of the thresholded GEVP [Eq.~\eqref{eq:gevp}] that belongs to the root. In the noisy runs, $\mathbf c$ comes from the noisy GEVP and the $\ket{\phi_n}$ are the exact Krylov vectors. The weight of the root on the exact eigenstate $\ket{E_m}$ is $p_m=|\langle E_m|\Psi\rangle|^2/\langle\Psi|\Psi\rangle$, with $\langle E_m|\Psi\rangle=\langle E_m|\phi_0\rangle\sum_n c_ne^{-inE_m\tau}$. We call a state resolved when some root has at least $90\%$ of its weight on it ($p_m\geq0.9$), and we call a root mixed when its weight on every state is below $90\%$.

Of the 443 points in Fig.~\ref{fig:noiseless_validation}, the 237 within chemical accuracy each come from a root with at least $97\%$ of its weight on the correct state. The other 206 have three causes. (i) In 172 cases, the state is not resolved. In 142 of them, it shares a root with a neighboring targeted state: the root with the largest weight on the state also has the largest weight on its neighbor. For example, at $1.7\,\tau_{\mathrm{alias}}$, $E_3$ and $E_4$ of BeH$_2$, which are $6$~mHa apart, share one root with $31\%$ of its weight on $E_3$ and $69\%$ on $E_4$. The two states then give one energy instead of two, so the estimates above $E_3$ shift by one state: $E_3$ is estimated by the shared root (error $4.7$~mHa), $E_4$ by the root of $E_5$ ($0.02$~Ha), and $E_5$ by the root of a higher, nontargeted state ($0.17$~Ha). The estimate of an unresolved state is thus a mixed root or the root of another state. The largest such error, $4.1$~Ha, occurs for N$_2$ at $0.3\,\tau_{\mathrm{alias}}$, where the roots mix many states. (ii) In 8 cases, like $E_5$ in this example, the state is resolved, but only one root is reported for a lower pair of unresolved states, so the root of the state is reported as the estimate of the next lower state, and its own estimate comes from a root with at least $87\%$ of its weight on a higher, nontargeted state. These errors are $0.16$--$0.19$~Ha. (iii) In the remaining 26 cases, the reported root has at least $90\%$ of its weight on the correct state, but its error is still $1.7$--$27$~mHa. An error above $30$~mHa therefore always means that the state or a lower one is not resolved. The missing points, for the two highest targets of N$_2$ at $0.2\,\tau_{\mathrm{alias}}$, its highest target at $0.3$ and $0.5\,\tau_{\mathrm{alias}}$, and the highest target of H$_2$O at $0.2$ and $0.3\,\tau_{\mathrm{alias}}$, all belong to states that are not resolved. No resolved state loses its root in validation, the window, or the acceptance level, and each of the 280 reported roots with at least $90\%$ of its weight on one state is unwrapped to within $27$~mHa of the energy of that state. This is much less than $\pi/\tau\geq1.76$~Ha, so every winding number is correct.

In the noisy runs of Fig.~\ref{fig:noisy_necessity}, the reported ground-state root has its largest weight on $E_0$ in 294 of the 296 realizations. In the other two, for BeH$_2$ at $\tau_{\mathrm{alias}}$, a root with its largest weight on $E_1$ lies $9$ and $65$~mHa below $E_0$ and is reported as the ground state by both methods. At the N$_2$ time steps where the base run alone fails, the accepted phantom peak is an image of a root whose largest weight is on a state $1.6$--$3.9$~Ha above $E_1$.
\end{document}